\documentclass[lettersize,journal]{IEEEtran}
\usepackage{amsmath}
\usepackage{amsmath,amsfonts}
\usepackage{algorithmic}
\usepackage{algorithm}
\usepackage{empheq}
\usepackage{array}
\usepackage[caption=false,font=footnotesize,labelfont=rm,textfont=rm]{subfig}
\usepackage{textcomp}
\usepackage{stfloats}
\usepackage{xcolor}
\usepackage{url}
\usepackage{verbatim}
\usepackage{graphicx}
\usepackage{cite}
\usepackage[T1]{fontenc}
\usepackage{booktabs,multirow}
\usepackage{pifont}
\usepackage{CJKutf8}
\usepackage{lipsum}
\usepackage{hyperref} 
\hypersetup{hidelinks,
  colorlinks=ture,
  linkcolor=black}

\begin{document}
\begin{CJK}{UTF8}{gbsn}

\title{Joint ADS-B in B5G for Hierarchical UAV Networks: Performance Analysis and MEC Based Optimization}

\author{Chao Dong, \IEEEmembership{Senior Member,~IEEE,} Yiyang Liao, Ziye Jia, \IEEEmembership{Member,~IEEE,}\\  
Qihui Wu, \IEEEmembership{Fellow,~IEEE}, and Lei Zhang

\thanks{Chao Dong, Yiyang Liao, Qihui Wu and Lei Zhang are with the College of Electronic and Information Engineering, Nanjing University of Aeronautics and Astronautics,
Nanjing 211106, China (e-mail: dch@nuaa.edu.cn; liaoyiyang@nuaa.edu.cn; wuqihui@nuaa.edu.cn; Zhang\_lei@nuaa.edu.cn).}
\thanks{Ziye Jia is with the College of Electronic and Information Engineering, Nanjing University of Aeronautics and Astronautics, Nanjing 211106, China, and also with the National Mobile Communications Research Laboratory, Southeast University, Nanjing, Jiangsu, 211111, China (e-mail: jiaziye@nuaa.edu.cn).
}}
\maketitle
\pagestyle{empty}  
\thispagestyle{empty}
\vspace{-1cm} 
\begin{abstract}
        Unmanned aerial vehicles (UAVs) play significant roles in multiple fields, which brings great challenges for the airspace safety. 
        In order to achieve efficient surveillance and break the limitation of application scenarios caused by single communication, we propose the collaborative surveillance model for hierarchical UAVs based on the cooperation of automatic dependent surveillance-broadcast (ADS-B) and 5G. Specifically, UAVs are hierarchical deployed, with the low-altitude central UAV equipped with the 5G module, and the high-altitude central UAV with ADS-B, which helps automatically broadcast the flight information to surrounding aircraft and ground stations. 
        Firstly, we build the framework, derive the analytic expression, and analyze the channel performance of both air-to-ground (A2G) and air-to-air (A2A). Then, since the redundancy or information loss during transmission aggravates the monitoring performance, the mobile edge computing (MEC) based on-board processing algorithm is proposed. Finally, the performances of the proposed model and algorithm are verified through both simulations and experiments. In detail, the redundant data filtered out by the proposed algorithm accounts for 53.48\%, and the supplementary data accounts for 16.42\% of the optimized data. This work designs a UAV monitoring framework and proposes an algorithm to enhance the observability of trajectory surveillance, which helps improve the airspace safety and enhance the air traffic flow management.
       
\end{abstract}

\vspace{-0.4cm} 
\section{Introduction}
\IEEEPARstart{W}{ith} the rapid development of aviation technologies, unmanned aerial vehicles (UAVs) are widely popularized. Due to the reliability, flexibility, adaptability, and high efficiency, UAVs are extensively applied in many fields, including the aerial photography, urban logistics, environmental monitoring, and UAV-assisted vehicular networks\cite{ref1}. However, as the number of UAVs increases, the flight safety becomes a non-negligible problem. Hence, it is necessary to strengthen the flight control and safety supervision to guarantee that the flight operation of UAVs does not affect the public safety and personal privacy\cite{ref2}. To deal with such problems, we consider joint the  automatic dependent surveillance-broadcast (ADS-B) technique \cite{ref3} in 5G, to realize the collaborative surveillance in the beyond 5G (B5G) networks.

\par The ADS-B system is composed of multiple ground stations (GSs) and airborne stations, which works at a specific frequency band, i.e., 1090MHz. Furthermore, the main service information of ADS-B includes the aircraft position, aircraft identification, aircraft velocity, and flight direction. In detail, an aircraft equipped with ADS-B can automatically broadcast its flight information to the nearby aircraft and GSs, which is conducive to the flight safety and air traffic management\cite{ref4}. However, due to the limited frequency band, excessive UAVs equipped with ADS-B interfere in the surveillance of GS towards civil planes. The typical impact is intensifying the collision of ADS-B packets, which leads to packets loss, and impairs the monitoring performance of GS towards civil planes\cite{ref5}. In order to weaken the hindrance on monitoring performance, ensure timely acquisition of the flight information of UAVs, and expand the monitoring capacity of GS, we consider the cooperation of ADS-B with 5G. 
\par Leveraging 5G to establish the network connection has the characteristics of ultra-high speed, ultra-low delay and ultra-large bandwidth\cite{ref6}, which meets the requirements for the wireless communication performance for UAV control and flight missions\cite{ref7}. UAVs equipped with 5G modules achieve real-time interactions between UAVs and GSs\cite{ref8}. However, single utilization of 5G brings the difficulties in the long distance transmission, and the integration of the network and air traffic management system of civil aviation. Therefore, we propose to deploy 5G modules on UAVs, by cooperating with ADS-B, to help GS acquire timely flight information. Besides, the central UAVs are equipped with computing resources, to handle the problem of trajectory optimization.
\par The deployment of UAVs adopts a hierarchical management structure\cite{ref9}. Besides, the mobile edge computing (MEC) modules are equipped on the central UAVs. In particular, the flight information of sub-UAVs is initially relayed to the central UAVs, and then transmitted to GS after MEC based processing. Due to the hierarchical structure, there exist two communication channels, i.e., the air-to-air (A2A) channel and air-to-ground (A2G) channel. To be specific, the A2A channel is established between the central UAV and sub-UAVs, while the A2G channel provides connections between the central UAV and GS. 
\par The challenges lie in how to construct the hierarchical monitoring surveillance for UAVs, carry out the channel performance analyze, and handle the problem of trajectory optimization. Hence, the contributions of this paper are summarized as follows:

\begin{itemize}
\item A cooperation framework of ADS-B and 5G for hierarchical UAV networks is designed to achieve the UAV flight information, aiming to improve the airspace safety and enhance the air traffic flow management.
\item We establish the A2A channel and A2G channel by stochastic modeling and deterministic modeling, respectively. Besides, we carry out the relevant analysis.
\item We propose the algorithm of on-board processing based on MEC, which can abandon redundant packets and supply missed packets, to increase the observability.
\item Extensive simulations are conducted to verify the influence of the height, density, path loss and transmitting power on the performance of the proposed network. Besides, we conduct real experiments to collect position packets for algorithm verification.
\end{itemize}

\par The rest of this paper is organized as follows.  Section \ref {S2} elaborates the related works. Besides, the network system model is introduced in Section \ref {S3}. Moreover, the stochastic performance analysis is presented in Section \ref {S4}. Additionally, Section \ref {S5} describes the mechanisms of on-board processing based on MEC. Section \ref {S6} provides the simulation results and corresponding analyses. Finally, Section \ref {S7} draws the conclusions.
\vspace{-0.2cm} 
\section{Related works}\label{S2}

\subsection{ADS-B, 5G, and MEC}
There exist some works about UAV equipped with ADS-B or 5G. For example, Ref. \cite{ref4} designs a system for UAV surveillance based on ADS-B and proposes related algorithms to handle trajectory planning. Ref. \cite{ref6} analyzes the impact of UAVs equipped with ADS-B on the civil planes at the frequency of 1090MHz, verifying the access capacity of the network layer.
Ref. \cite{ref8} points out that 5G fully meets the requirements for the wireless communication performance of UAVs flight control and missions, achieving real-time interaction between UAVs and GS. Ref. \cite{ref10} develops a novel trust-based security scheme for 5G UAV communication systems, improving the communication performance and evaluating the reliability. Besides, there exist some related works about the deployment of the MEC on UAVs, aiming to improve the performance of UAV communications by appropriately utilizing computing resources at the edge of networks\cite{ref11}. For example, Ref. \cite{ref12} considers a task offloading problem for a UAV-assisted MEC system, and designs an integrated cloud-edge network, thus minimizing the weighted cost of latency and energy consumption. Ref. \cite{ref13} proposes a joint communication and computation optimization network model of UAV swarms, which leverages MEC to decrease the response delay and increase the efficiency of network resources utilization. Ref. \cite{ref14} designs a UAV-assisted MEC system to verify the UAV trajectory optimization, resource allocation, and tasks offloading. The work handles the problem of minimizing the energy consumption of mobile devices and computing power of UAVs. Ref. \cite{ref15} utilizes the resources of computation and storage on UAVs to propose an elastic collaborative MEC intelligence, enhancing the network instability.

\begin{figure}[t]
        \centerline{\includegraphics[width=0.85\linewidth]{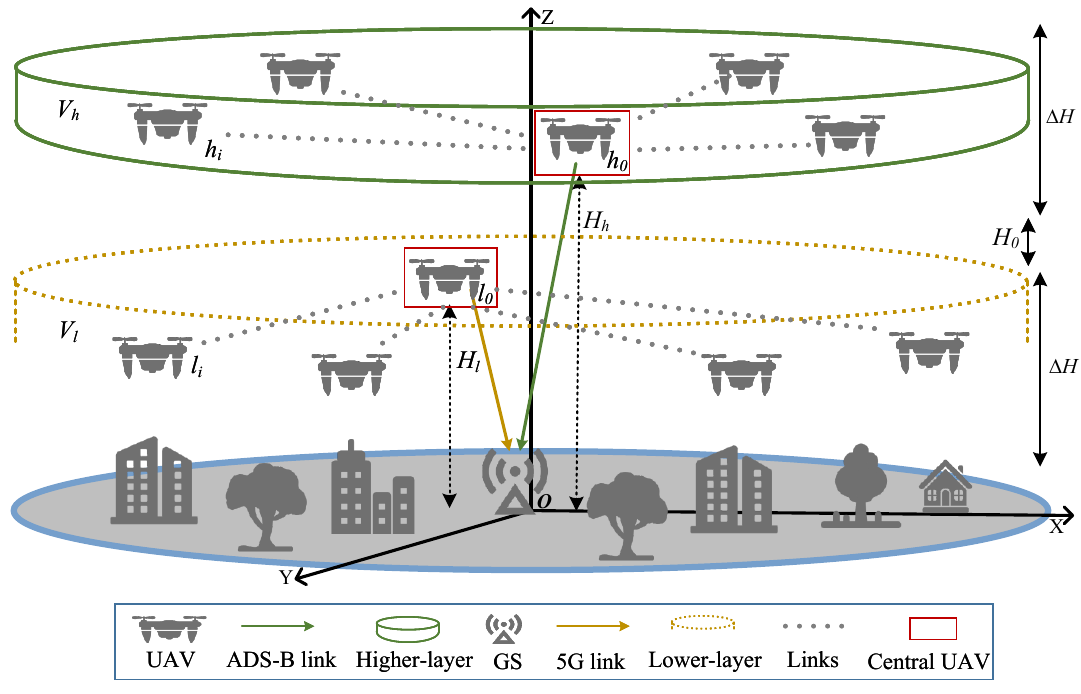}}
        \caption{Hierarchical UAV network model.}
        \label{f1}
\end{figure}

\subsection{Channel Modeling}

\par Additionally, there exist some works involving the establishment of channel models related with UAVs. For example, Ref. \cite{ref16} points out that channel modeling has two methods, including deterministic modeling and stochastic modeling. The deterministic modeling has significant precision but lacks universality. On the contrary, the stochastic modeling is flexible but lacks pertinence. Ref. \cite{ref17} employs a deterministic channel model that simply abstracts the physical layer, and effectively captures the effects of channel strength, broadcast and superposition in wireless channels. Ref. \cite{ref18} leverages deterministic model to investigate the optimization for massive multi-input multi-output near-field wireless communications in respect to line-of-sight (LOS) channel. Ref. \cite{ref19} states that UAV wireless networks have natural spatial random characteristics and the channel has fading and shadowing characteristics. Hence, the stochastic modeling is based on the theory of the stochastic geometry (SG). Ref. \cite{ref20} leverages the Poisson point process (PPP) in SG to simulate the distribution of the coexistence of UAVs and planes, aiming to analyze the signal interference. The authors in Ref. \cite{ref13} utilize SG to analyze the response delay and the successful transmission probability of the UAV network, and deduce the analytical form of signal-to-interference-plus-noise ratio (SINR).
\subsection{State of the Art}
\par As far as the authors know, apart from ADS-B and 5G, there are other ways to establish the communication, so as to achieve timely surveillance. Ref. \cite{ref21} points out that, based on Bluetooth or WiFi, Remote ID is a broadcast technology that offers the identification and position information of UAVs\cite{ref22}. Besides, Ad-hoc is a kind of temporary network that does not rely on traditional infrastructure\cite{ref23}. It directly establishes communication with peers\cite{ref24}. Moreover, incorporating UAVs with satellite communication systems overcomes the geographical limitations\cite{ref25}. However, all the demands faced by the aforementioned technologies can be met by ADS-B and 5G, which cover the long-distance, short-distance, high-speed and low-latency communications. Therefore, we innovatively utilize ADS-B in B5G to build the hierarchical UAV networks, which guarantees the integrated management by civil aviation authorities, and achieves various applications in different scenarios, aiming to propose a safe collaborative surveillance with respect to hierarchical UAV networks.

\par Traditional trajectory optimization involves two major problems, i.e., trajectory planning and trajectory prediction, and mainly adopts methods such as particle filtering\cite{ref26}, Kalman filtering\cite{ref27}, and neural networks\cite{ref28}. However, in this paper, the trajectory optimization focuses on the problem of monitoring observability, which values timeliness over accuracy. Therefore, we designed the on-board processing algorithm, inspired by the sliding window filtering, to improve the observability and solve a practical engineering problem.

\section{System Model}\label{S3}

\begin{table}[t]
        \scriptsize
        \caption{Key parameters} 
        \centering
        \begin{tabular}{ll} 
                \toprule 
                \textbf{Notation} & \textbf{Definition} \\
                \midrule 
                $\lambda_l$/$\lambda_h$ & The density of low/high-altitude UAVs \\
                $V_l$/$V_h$ & The low/high-altitude airspace\\
                $l_0$/$h_0$ & The central UAV of low/high-altitude airspace\\
                $l_j$/$h_i$ & The sub-UAVs of low/high-altitude airspace\\
                $H_l$/$H_h$ & The height of $l_0$/$h_0$\\
                $d_{l_j}$/$d_{h_i}$ & The distance between $l_j$/$h_i$ and $l_0$/$h_0$\\
                $\Delta H$ & The thickness of the airspace\\
                $P_{cl}$/$P_{ch}$ & The transmitting power of $l_0$/$h_0$\\
                $P_s$ & The transmitting power of all sub-UAVs\\
                $G_a$/$G_g$ & The total gain of A2A/A2G channel\\
                $\Delta \varphi$ & The difference of phase between paths \\
                $\psi$ & The grazing angle of signal\\
                $\mathcal{D}$ & The divergence factor\\
                $\Gamma_\perp $/$\Gamma_t $ & The reflection coefficient without/with the \\
                &influence of $\mathcal{D}$\\
                $|\Gamma_t|$/$\phi_t$ & The length/phase of $\Gamma_t $\\
                $\lambda_{5G}$/$\lambda_{ADS-B}$ & The wavelength of 5G/ADS-B signal\\
                $f_{5G}$/$f_{ADS-B}$ & The frequency of 5G/ADS-B signal\\
                $B_{5G}$/$B_{ADS-B}$ & The bandwidth of 5G/ADS-B signal\\
                $\varepsilon_0$/$\varepsilon_r$ & The dielectric constant/The relative dielectric constant\\
                $\sigma$ & The electric conductivity\\
                $PL$ & The path loss of A2G channel (dB)\\
                $\varpi$ & SNR between the central UAV and GS\\
                $N_0$/$n_0$ & Gaussian noise/The power density of noise \\
                $\delta$ & The index of path loss in A2A\\
                $\rho$ & The index of small scale fading in A2A \\
                $\gamma$ & SINR between the central UAV and sub-UAV \\
                $\theta_l$/$\theta_h$ & The received threshold of $l_0$/$h_0$\\
                $P_{cov}$ & The coverage probability\\
                $I_{l_j}^g$/$I_{h_i}^k$ & The $g$-$th$/$k$-$th$ 5G/ADS-B packet from the \\
                &$j$-$th$/$i$-$th$ sub-UAV\\
                $Lon_{h_i}^k$/$Lat_{h_i}^k$/$Alt_{h_i}^k$ & The longitude/latitude/altitude information of $I_{h_i}^k$\\
                $\mathcal M_{h_i}$ & The vector of Minkowski for $h_i$\\
                $m_{h_i}^n$ & The $n$-$th$ Minkowski distance in $\mathcal M_{h_i}$\\
                $X$/$B$ & The coefficient/constant determinant\\
                \bottomrule 
        \end{tabular}
        \label{tab1}   
\end{table}

 Fig. \ref{f1} depicts the hierarchical UAV network model. In detail, a group of UAVs with density $\lambda_l$ are randomly distributed in the low-altitude airspace $V_l$, corresponding to the orange dotted line area. Another group of UAVs with density $\lambda_h$ scatter in the high-altitude airspace $V_h$, as the green solid line area shows. Besides, a central UAV exists in each layer. The low-altitude central UAV is equipped with 5G module, while the high-altitude central UAV transmits the flight information through the ADS-B system. $H_h$ and $H_l$ are regarded as the altitude of the central UAV $h_0$ and central UAV $l_0$, respectively. $\Delta H$ is the height of both airspaces. Furthermore, there is an isolation layer with height $H_0$ between the two airspaces, aiming to reduce the interference among the UAVs from different airspaces. It is assumed that there is a GS located at \textsl{\textbf{O}}(0, 0, 0) in the space. GS can receive the 5G flight information from $l_0$ and the ADS-B flight information from $h_0$. Additionally, since the central UAVs serve as aerial stations, which are quasi-static, we establish the A2G channels by utilizing deterministic models, which have the merits of strong pertinence, precise fit and high accuracy. Besides, considering the mobility and maneuverability of sub-UAVs, we establish the A2A channels by leveraging stochastic models, which have the advantages of universality and flexibility. Key parameters and corresponding definitions are summarized in TABLE \ref {tab1}.

 \vspace{-0.2cm} 
 \subsection{Data Format}
 \subsubsection{Data Format of ADS-B}
 \par
\begin{figure}[t]
        \centerline{\includegraphics[width=1\linewidth]{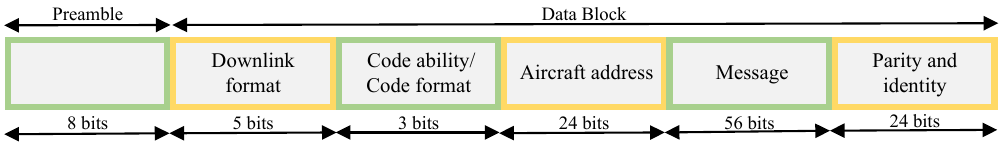}}
        \caption{Structure of ADS-B.}
        \label{f2}
\end{figure}

 As shown in Fig. \ref{f2}, a single ADS-B data consists of an 8-bit preamble and a 112-bit data block\cite{ref29}. The down link format (DF) is 5 bits long, which is used to distinguish the type of transponder. If DF=17, the third field is the code ability. If DF=18, the third field is the code format. Besides, the aircraft address field is 24 bits long. The message field contains the service information. Finally, the length of parity and identity field is 24 bits\cite{ref30}.

\begin{figure}[t]
        \centerline{\includegraphics[width=1\linewidth]{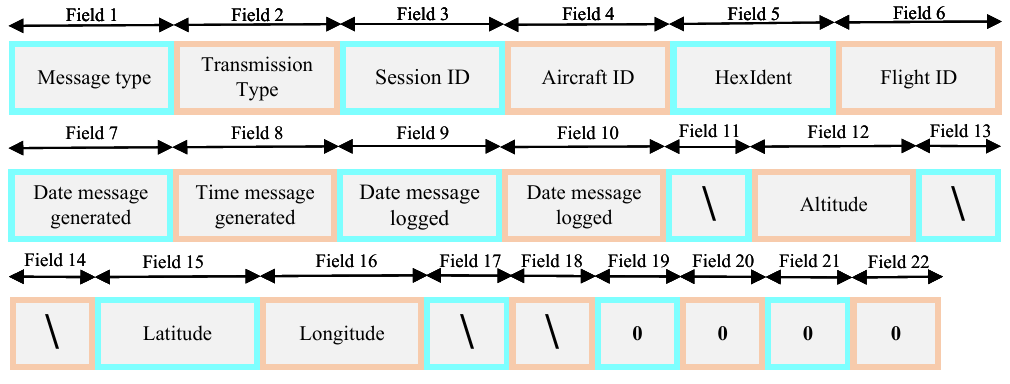}}
        \caption{Structure of airborne position TM.}
        \label{f3}
\end{figure}
 \subsubsection{Data Format of 5G}
 \par To facilitate the information processing of GS, we transmit the Kinetic Avionics data via the 5G network, and the specific data format is compatible with ADS-B. In this work, we primarily focus on the longitude, latitude and altitude information. Therefore, we utilize the airborne position message, illustrated in Fig. \ref{f3}. Specifically, field 1 is the fixed character, i.e., "MSG". The transmission type is set as "3", corresponding to the airborne position message. Fields 3 and 4 indicate the database session record number and database aircraft record number, respectively. Besides, field 5 represents the aircraft mode S hexadecimal code and field 6 symbolizes the database flight record number. The meanings of other fields are shown in the Fig. \ref{f3}. Moreover, a slash denotes that the field is empty, and the last four fields are fixed as "0" in respect to the airborne position message\cite{ref31}.
 \vspace{-0.3cm} 
 \subsection{Parameters of UAV Model}
\par Denote the set of high-altitude UAVs as $\mathcal H=\{h_0,h_1,...,h_i,...,h_u\}$ and $i\in (1,u)$. Indicate the set of low-altitude UAVs as $\mathcal L=\{l_0,l_1,...,l_j,...,l_v\}$ and $j\in (1,v)$. $u$ and $v$ represent the number of the corresponding sub-UAVs. In detail, $h_0$ in $\mathcal H$ symbolizes the central UAV in the high-altitude airspace while $l_0$ in $\mathcal L$ indicates the central UAV in the low-altitude airspace. Additionally, the coordinate of the $i$-$th$ UAV in set $\mathcal H$ is $(x_{h_i}$, $y_{h_i}$, $z_{h_i})$ and the coordinate of the $j$-$th$ UAV in set $\mathcal L$ is $(x_{l_j}$, $y_{l_j}$, $z_{l_j})$. The X-axis coordinates for all UAVs range within $[-L_x, L_x]$, the Y-axis coordinates range within $[-L_y, L_y]$, and the Z-axis coordinates are from $[0, L_z]$. In specific, $L_z=2\Delta H+H_0$. The euclidean distance between UAV $h_i$ and $h_0$ is $d_{h_i}=\sqrt{(x_{h_i}-x_{h_0})^2+(y_{h_i}-y_{h_0})^2+(z_{h_i}-z_{h_0})^2}$, and the euclidean distance between UAV $l_j$ and $l_0$ is $d_{l_j}=\sqrt{(x_{l_j}-x_{l_0})^2+(y_{l_j}-y_{l_0})^2+(z_{l_j}-z_{l_0})^2}$.  $P_s$ represents the transmitting power of the sub-UAV. In addition, $P_c$ declares the transmitting power of the central UAV, which is further divided into $P_{ch}$ of $h_0$ and $P_{cl}$ of $l_0$, respectively. $G_g$ represents the total gain of the A2G channel, including the transmitter gain in the central UAV and receiver gain at GS. Besides, $G_a$ represents the total gain of the A2A channel, including the transmitter gain in sub-UAV and the receiver gain in the central UAV. Besides, $P_s$, $G_g$ and $G_a$ are same for all links in the model. 
 
\subsection{A2G Channel Model} 
\begin{figure}[t]
        \centerline{\includegraphics[width=0.8\linewidth]{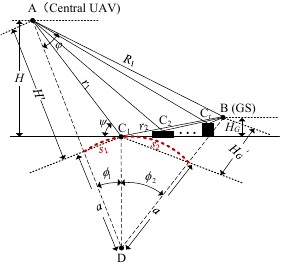}}
        \caption{A2G channel.}
        \label{f4}
\end{figure}       
 In the case of long link distances, the shape of earth is approximately equivalent to a sphere of radius $a$\cite{ref32}. Therefore, we model the A2G into a curved-earth multi-rays (CEMR) model, as shown in Fig. \ref{f4}. Moreover, the black rectangles represent the reflectors. The distance of LoS component between A and B is $R_1$. Additionally, we take the non-line-of-sight (NLoS) component passing ${\rm C_1}$ for example, whose distance is $R_2$, and $R_2$=$r_1$+$r_2$. Furthermore, D is regarded as the center of earth and ${\rm C}_t$ is supposed to be the $t$-$th$ reflection point. Initially, we denote the distance between the central UAV and the surface as $H$, and assume the distance between the central UAV and the tangential plane at ${\rm C_1}$ as $H'$. Specifically, when concerning the high-altitude airspace, $H$ refers to $H_h$. Besides, $H$ conducts as $H_l$ when involving the low-altitude airspace. Subsequently, $H_G$ expresses the distance between GS and the surface. $H_G'$ declares the distance between GS and the tangential plane at ${\rm C_1}$. Since we arrange the central UAVs to hover in a small area above the GS, the horizontal distance is ignored in comparison with the vertical distance, i.e., $H\approx H'$ and $H_G\approx H_G'$. In triangle ABD, according to the law of cosine\cite{ref33}, we have
\begin{equation} 
        R_1^2=(a+H')^2+(a+H_G')^2-2(a+H')(a+H_G')\cos(\phi),
        \label{eq1}
\end{equation}
where $\phi$=$\phi_1$+$\phi_2$, $s$=$s_1$+$s_2$=$a\phi$, AB=$R_1$, AD=$H'$+$a$, and BD=$H_G'$+$a$. By substituting the above formulas, three intermediate variables are defined, which are conducive to the simplified exhibition of equations\cite{ref34}. In detail, three intermediate variables $\omega_1$, $\omega_2$, and $\omega_3$ are indroduced as
\begin{equation}
        \begin{cases} 
        \omega_1=\frac{s^2}{4a(H'+H_G')}, \\ 
        \omega_2=\frac{H'-H_G'}{H'+H_G'}, \\
        \omega_3=2\sqrt{\frac{\omega_1+1}{3\omega_1}}\cos\bigg\{\frac{\pi}{3}+\frac{1}{3}\arccos\bigg[\frac{3\omega_2}{2}\sqrt{\frac{3\omega_1}{(\omega_1+1)^3}}\bigg]\bigg\}. 
        \label{eq2}
        \end{cases} 
\end{equation}
By combining Eq. (\ref{eq1}), (\ref{eq2}) and $s$=$a\phi$, we obtain
\begin{equation}
        \begin{cases}
        s_1=\frac{s(1+\omega_3)}{2}, \\ 
        s_2=s-s_1, \\
        \phi_1=\frac{s_1}{a}.
        \end{cases} 
\end{equation}
The phase difference of the signal between path AB and path A${\rm C_1}$B is $\Delta \varphi_1$, calculated as 
\begin{equation}
        \Delta \varphi_1=\frac{2\pi\Delta s}{\lambda}, 
\end{equation}
where $\Delta s$=$2s_1s_2\psi^2/s$, $\psi$=$(H'+H_G')[1-\omega_1(1+\omega_2^2)]/s$ and $\lambda$ is the wavelength. In detail, $\lambda$=$c/f$, in which $c$ is the velocity of light and $f$ is the frequency. Additionally, we assume that the reflection coefficient of earth is $\Gamma_\perp $. Since we consider all UAVs are equipped with vertically polarized antennas, $\Gamma_\perp$ is simplified as
\begin{equation} 
        \Gamma_\perp=\frac{(\varepsilon_r-jb)\sin\psi-\sqrt{(\varepsilon_r-jb)-\cos\psi} }{(\varepsilon_r-jb)\sin\psi+\sqrt{(\varepsilon_r-jb)-\cos\psi}}, 
\end{equation}
where $\varepsilon_r$ is the relative dielectric constant and $j^2$=$-1$. Besides, $b$=$\sigma/(2\pi f\varepsilon_0)$, where $\sigma$ is the electric conductivity, and $\varepsilon_0$ is the dielectric constant.
Nevertheless, the reflection coefficient of point ${\rm C_1}$ is $\Gamma_{1}$=$|\Gamma_{1}|e^{j\phi_{1}}$=$\mathcal{D} \Gamma_\perp$, where $\mathcal{D}$ is the divergence factor. Further, $\mathcal{D}$ is defined as
\begin{equation}
  \mathcal{D}=\bigg[1+\frac{2r_1r_2}{a(r_1+r_2)\sin\psi }\bigg]^{-\frac{1}{2}},
\end{equation}
where $r_1$ is obtained from the law of cosine in triangle A${\rm C_1}$D, i.e., ${r_1}^2$=$(a$+$H')^2$+$a^2$-$2a(a$+$H')\cos\phi_1$ and $r_2$ is obtained from the law of cosine in triangle ${\rm C_1}$BD, i.e., ${r_2}^2$=$(a$+$H_G')^2$+$a^2$-$2a(a$+$H_G')\cos\phi_2$. $E_{LoS}$ and $E_{NLoSt}$ respectively symbolize the signal strength of the LoS path and the $t$-$th$ NLoS path, and $|E_{LoS}|^2$=$P_cG_t$, where $G_t$ denotes the transmitter gain of the central UAV. Hence, the signal strength $E_G$ received by GS is the vector sum of the signals at point B, whose analytic form is 
\begin{align}
        E_G&=E_{LoS}+\sum_{t = 1}^{\lfloor\frac{\pi}{2\varrho }\rfloor}  E_{NLoSt} \nonumber\\ 
        &=E_{LoS}\bigg[1+\sum_{t = 1}^{\lfloor\frac{\pi}{2\varrho }\rfloor}  |\Gamma_t|e^{-j(\Delta \varphi_t-\phi_t)}\bigg]\nonumber\\ 
        &=E_{LoS}\bigg\{\bigg[1+\sum_{t = 1}^{\lfloor\frac{\pi}{2\varrho }\rfloor}|\Gamma_t|\cos(\Delta \varphi_t-\phi_t)\bigg]\nonumber\\
        &-j\sum_{t = 1}^{\lfloor\frac{\pi}{2\varrho }\rfloor}|\Gamma_t|\sin(\Delta \varphi_t-\phi_t)\bigg\},
\end{align}
where $\varrho$ denotes the beamwidth.
Moreover, the power $P_G$ of the signal received by GS is formulated as
\begin{equation}
        P_G=\frac{\|E_G\|^2G_r\lambda^2}{(4\pi R_1)^2}, 
\end{equation}
where $G_r$ is the receiver gain at GS and $G_tG_r$=$G_g$. Hence, the path loss of the A2G communication system is defined as
\begin{equation}
        PL=-10{\rm log}_{10}\bigg(\frac{P_G}{P_c}\bigg). 
        \label{eq10}
\end{equation}
Additionally, $\varpi$ denotes the signal-to-noise ratio (SNR), i.e.,
\begin{equation}
        \varpi=10{\rm log}_{10}(P_G)-10{\rm log}_{10}(n_0B),
        \label{eq11}
\end{equation}
where $n_0$ is the noise power density and $B$ is the bandwidth.
\subsection{A2A Channel Model}
 Considering the hierarchical airspace, we set up an isolation layer between the airspaces, and only the interference in the same airspace is considered in the model. The sub-UAVs in the same airspace communicate with the central UAV. Considering the high-altitude airspace, the path loss from the $i$-$th$ sub-UAV to the central UAV is proportional to $d_{h_i}^{-\delta}$, where $d_{h_i}$ denotes the distance between the sub-UAV and central UAV. $\delta$ indicates the path loss index. Besides, $\rho_{h_i}$ follows an exponential distribution with mean value of 1, indicating the gain of small scale fading channel. Similarly, Gaussian noise $N_0$ is added to the model, i.e., $N_0$=$n_0B$. We leverage $\gamma$ to represent SINR. $\gamma_{h_i}$, the desired signal sent by the $i$-$th$ sub-UAV $h_i$ in set $\mathcal H$ to the central UAV $h_0$, is calculated as

\begin{equation} 
  \gamma_{h_i}=\frac{{P_s}{G_a}{\rho_{h_i}}{d_{h_i}^{-\delta}}}{N_0+{P_s}S_{\mathcal {H}\backslash \{{h_i}\}}},
\end{equation}

\noindent where

\begin{equation} 
S_{\mathcal{H}\backslash \{{h_i}\}}=\sum_{{h}\in \mathcal{H}\backslash \{{h_i}\}}{G_a}{\rho_{h_i}}{d_{h_i}^{-\delta}}.
\label{eq12}
\end{equation}

\noindent Similarly, as for the low-altitude airspace, SINR $\gamma_{l_j}$ of the desired signal sent by the $j$-$th$ sub-UAV $l_j$ in set $\mathcal L$ to the central UAV $l_0$ is  
\begin{equation} 
        \gamma_{l_j}=\frac{{P_s}{G_a}{\rho_{l_j}}{d_{l_j}^{-\delta}}}{N_0+{P_s}S_{\mathcal {L}\backslash \{{l_j}\}}},
      \end{equation}
      
\noindent in which 
\begin{equation} 
      S_{\mathcal{L}\backslash \{{l_j}\}}=\sum_{{l}\in \mathcal{L}\backslash \{{l_j}\}}{G_a}{\rho_{l_j}}{d_{l_j}^{-\delta}}.
\end{equation}

\begin{figure*}[t] 
        \centering
        \begin{minipage}{1\linewidth}
        \centering
        \subfloat[Packet abandonment.]{
        \includegraphics[width=0.45\linewidth]{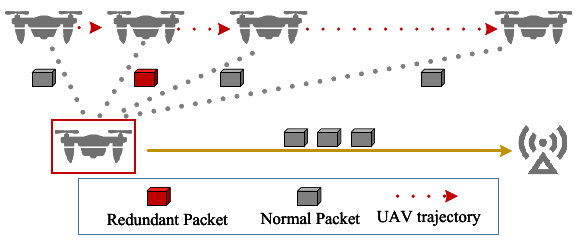}
        \label{f5.1}}
        \subfloat[Packet supplement.]{
        \includegraphics[width=0.45\linewidth]{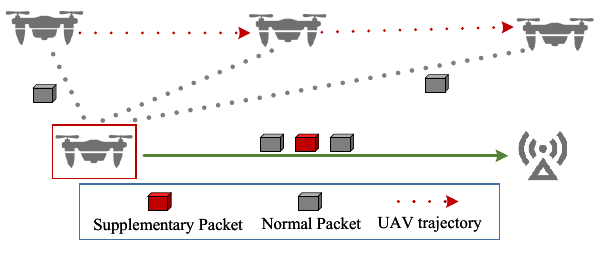}
        \label{f5.2}}
        \caption{On-board processing.}
        \label{f5}
        \end{minipage}
\end{figure*}

\section{Analysis of Stochastic Geometry}\label{S4}
In this section, we use the PPP in SG theory to fit the distributions of sub-UAVs in the A2A. Besides, the analytic function is derived for the performance analysis. Taking the high-altitude airspace $V_h$ for example, all sub-UAVs follow the nearest neighbor association strategy\cite{ref35}, i.e.,

\begin{equation} 
  P\{d_{h}> R^*\}={\rm exp}(-λ\lambda_h V_h)={\rm exp}\bigg(-\frac{4}{3}\pi\lambda_h{d^3_{h}}\bigg),
\end{equation}

\noindent where $d_{h}$$\geq$0, and $R^*$ is the longest distance that the central UAV can serve. Therefore, the cumulative distribution function (CDF) of the distance $d_{h}$ is
\begin{equation} 
  F(d_{h})= P\{d_{h}\leq R^*\}=1-{\rm exp}\bigg(-\frac{4}{3}\pi\lambda_h{d^3_{h}}\bigg),
\end{equation}

\noindent and the probability density function (PDF) of $d_{h}$ is a derivative of CDF, demonstrated as
\begin{equation} 
  f(d_{h})=4\pi\lambda_h{d^2_{h}}{\rm exp}\bigg(-\frac{4}{3}\pi\lambda_h{d^3_h}\bigg).
\end{equation}

Supposing the transmission will be succesful if the $\gamma_{h_i}$ is greater than the received threshold $\theta_h$  of the central UAV $h_0$, so the coverage probability of the $i$-$th$ sub-UAV in $\mathcal H$ is
\begin{equation} 
  P_{cov}^i=\mathbb{E}[P(\gamma_{h_i}\geq\theta_h|d_{h_i})].
  \label{eq19}
\end{equation} 
\noindent Since $\gamma_{h_i}$ is a function of $d_{h_i}$, $P_{cov}^i$ is further expressed as

\begin{equation}
        P_{cov}^i=\int_{0}^{\infty}P(\gamma_{h_i}\geq\theta_h|d_{h_i})f(d_h) \,d(d_h).
        \label{eq20}
\end{equation}

\par It is assumed that the average gain of small scale fading channel in the A2A is a random variable following the Gamma distribution with mean value of 1\cite{ref36}, which is depicted as

\begin{equation} 
  f(\rho)=\frac{\iota ^\iota    }{\Gamma(\iota )} \rho ^{\iota -1}e^{-\iota \rho }.
\end{equation}

\noindent When $\iota  $ is 1, the channel is considered as Rayleigh fading. $\rho $ follows an exponential distribution with mean value of 1. The PDF of $\rho $ is $f(x )$=$e^{-x}$, i.e., ${\rho _{h_i}}$$\sim$${\rm exp}(1)$ and ${\rho _{l_j}}$$\sim$${\rm exp}(1)$. Hence, $P(\gamma_{h_i}\geq\theta_h|d_{h_i})$ in Eq. (\ref{eq19}) is further deduced as 
\vspace{-0.1cm} 
\begin{align}
  P(\gamma_{h_i}\geq\theta_h|d_{h_i})&=P\bigg({\rho_{h_i}}\geq\frac{\theta_h{{d^\delta _{h_i}}}(N_0+{P_s}S_{\mathcal{H}\backslash \{{h_i}\}}) }{P_sG_a}\bigg)\nonumber\\
  &={\rm exp}\bigg(\frac{-\theta_h{{d^\delta_{h_i}}}(N_0+{P_s}S_{\mathcal{H}\backslash \{{h_i}\}})}{P_sG_a}\bigg)\nonumber\\
  &={\rm exp}\bigg(\frac{-\theta_h{{d^\delta _{h_i}}}N_0}{P_sG_a}\bigg)\mathbb{L}_{S_{\mathcal{H}\backslash \{{h_i}\}}}\bigg(\frac{\theta_h{{d^ \delta _{h_i}}}}{G_a}\bigg).
  \label{eq22}
\end{align}
\noindent Let $\frac{\theta_h{{d^\delta_{h_i}}}}{G_a}=\Lambda_{h_i} $, and we have
\begin{equation}
  \mathbb{L}_{S_{\mathcal{H}\backslash \{{h_i}\}}}\bigg(\frac{\theta_h{{d^\delta_{h_i}}}}{G_a}\bigg)=\mathbb{L}_{S_{\mathcal{H}\backslash\{{h_i}\}}}(\Lambda_{h_i} )=\mathbb{E}[e^{-\Lambda_{h_i}(S_{\mathcal{H}\backslash \{{h_i}\}})}],
  \label{eq23}
\end{equation}
\noindent which is the Laplace transform of $S_{\mathcal{H}\backslash \{{h_i}\}}$. By substituting Eq. (\ref{eq12}) into Eq. (\ref{eq23}), we further derive
\begin{align}
        \mathbb{L}_{S_{\mathcal{H}\backslash\{{h_i}\}}}&(\Lambda_{h_i})=\mathbb{E}\bigg[{\rm exp}\bigg(-\Lambda_{h_i}\sum_{{h}\in \mathcal{H}\backslash \{{h_i}\}}{G_a}{\rho _{h}}{d_{h}^{-\delta }}\bigg)\bigg]\nonumber\\
        &\overset{(\textbf{a})}{=}\mathbb{E}\bigg[\prod_{h\in \mathcal{H}\backslash\{{h_i}\}}\frac{1}{1+\Lambda_{h_i} G_ad_{h}^{-\delta }}\bigg]\nonumber\\
        &\overset{(\textbf{b})}{=}{\rm exp}\bigg[-\lambda_h\int_{V_h}\bigg(1-\frac{1}{1+\Lambda_{h_i} G_ad_{h}^{-\delta }}\bigg)d(d_{h})\bigg]\nonumber\\
        &\overset{(\textbf{c})}{=}{\rm exp}\bigg[-\lambda_h\int_{-L_x}^{L_x}\int_{-L_y}^{L_y}\int_{0}^{L_z}1-\nonumber\\
        &\quad\quad\frac{1}{1+\Lambda_{h_i} G_ad_{h}^{-\delta }}dxdydz\bigg]\nonumber\\
        &\overset{(\textbf{d})}{=}{\rm exp}(-\lambda_h\Theta_{h_i} ).
        \label{eq24}
      \end{align}
\noindent  Wherein, (\textbf{a}) is obtained by the moment generating function, based on THEOREM 4.9 in \cite{ref37}. (\textbf{b}) follows the probability generating function, based on eq. (90) in \cite{ref38}, which serves as a mathematical tool that converts the expectation of a continuous multiplication of functions into an integral over the PPP domain.
Besides, $d_{h}$ in (\textbf{b}) can be further expressed as $\sqrt{(x_{h}-x_{h_0})^2+(y_{h}-y_{h_0})^2+(z_{h}-z_{h_0})^2}$, demonstrated as (\textbf{c}).
Finally, $\Theta_{h_i}$ in (\textbf{d}) represents the triple integral in (\textbf{c}), aiming to simplify the formula.
\par \noindent By substituting Eq. (\ref{eq22}), (\ref{eq23}) and (\ref{eq24}) into (\ref{eq20}), $P_{cov}^i$ is calculated as 
\begin{align}
    P_{cov}^i&=\int_{0}^{\infty}4\pi\lambda_h{d^2_{h}}{\rm exp}\bigg(\frac{-\theta_h{{d^\delta_{h_i}}}N_0}{P_sG_a}\nonumber\\
    &-\lambda_h\Theta_{h_i}-\frac{4}{3}\pi\lambda_h{d^3_{h}}\bigg)d(d_{h}). 
\end{align}
The derivation processes of formulas aforementioned are also applicable to the calculations of coverage probability for sub-UAVs in low-altitude airspace $V_l$. Hence, the coverage probability of the $j$-$th$ sub-UAV in $\mathcal L$ is presented as
\begin{align}
        P_{cov}^j&=\int_{0}^{\infty}4\pi\lambda_l{d^2_{l}}{\rm exp}\bigg(\frac{-\theta_l{{d^\delta_{l_j}}}N_0}{P_sG_a}\nonumber\\
        &-\lambda_l\Theta_{l_j}-\frac{4}{3}\pi\lambda_l{d^3_{l}}\bigg)d(d_{l}). 
    \end{align}

\section{On-board processing based on MEC}\label{S5}
During the real transmission of flight information, there are two abnormal conditions:
\begin{itemize}
        \item Redundant flight packet: Packet of redundant information increases the calculation burden and causes trajectory overlap in the surveillance, further impairing the observability.
        \item Discontinuous flight packet: Packet loss due to error or collision leads to the discontinuity of trajectory, further impairing the observability.
\end{itemize}
\par If the trajectory optimization is solely entrusted to GS for completion, there exsits a high probability that the packet loss will be aggravated owing to the long transmission distance, which further impairs the optimization performance. Consequently, we deploy MEC on the central UAV and propose an on-board processing mechanism. As such, we address the aforementioned two issues and achieve the timely optimization.
\par In short, the optimization objective is to improve the observability of trajectory monitoring, in order to help air traffic controllers visualize the flight path intuitively.

\subsection{Packet Abandonment} Compared with the civil aircraft, UAV flies at a lower speed. Thus, UAV trajectories change slowly, so GS does not need to continuously receive position packets in short intervals. Meanwhile, abandoning redundant packets reduces the channel occupation of a single UAV, and improves the monitoring capacity of GS. As depicted in Fig. 5\subref{f5.1}, after receiveing two consecutive packets from the sub-UAV, the central UAV processes them. If the information of two packets is similar, the second packet is regarded as a redundant packet. Based on the on-board processing mechanism, the central UAV refuses to relay the redundant packet, which effectively prevents the overlap of trajectory on the surveillance and reduces the computing burden.
\subsection{Packet Supplement} Due to the effect of path loss and channel shadowing, the packet loss occurs during the transmission\cite{ref39}. Besides, on account of the limited processing capacity, the packet collision happens when multiple packets arrive at the central UAV simultaneously\cite{ref6}. Generally, packet loss and collision cause the central UAV to miss packets. As illustrated in Fig. 5\subref{f5.2}, after receiveing two consecutive packets from the sub-UAV, the central UAV processes them. If the information of two packets is disparate, the second packet is regarded as a discontinuous packet. Based on the on-board processing mechanism, the central UAV replenishs a supplementary packet, which ensures the continuous trajecorty of UAVs for the surveillance.
\subsection{On-board Processing Mechanism}
After receiving several position packets, GS generates the trajectory of the targeted UAV. When packets are continuously lost, GS clears the flight path of the UAV, and the UAV is deemed to leave the controlled airspace. For the flight surveillance, the position information is significant. We respectively leverage $Lon$, $Lat$ and $Alt$ to symbolize the longitude, latitude and altitude in a position packet. Hence, the position vector of the $i$-$th$ high-altitude UAVs in the $k$-$th$ packets can be classified as
\begin{equation}
        I_{h_i}^k=[Lon_{h_i}^k, Lat_{h_i}^k, Alt_{h_i}^k].
        \label{eq27}
\end{equation}
Similarly, the position vector of the $j$-$th$ low-altitude UAVs in the $g$-$th$ packets can be classified as
\begin{equation}
        I_{l_j}^g=[Lon_{l_j}^g, Lat_{l_j}^g, Alt_{l_j}^g].
\end{equation}
The following mechanisms can be applied to $I_{h_i}^k$ of UAV $h_i$ as well as $I_{l_j}^g$ of UAV $l_j$.
\begin{figure}[t]
        \centerline{\includegraphics[width=0.85\linewidth]{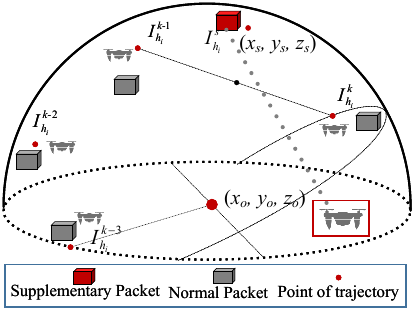}}
        \caption{Mechanism of supplement.}
        \label{f6}
\end{figure}
\subsubsection{Mechanism of Abandonment}
The central UAV records the first $N$ position packets of the ${i}$-$th$ sub-UAV and compute the Minkowski distances of adjacent position vectors in turns\cite{ref40}. The Minkowski vector $\mathcal M_{h_i}$ is defined as
\begin{equation}
        \mathcal M_{h_i}=\{m_{h_i}^1,m_{h_i}^2,...,m_{h_i}^{n},...,m_{h_i}^{N}\}.
        \label{eq29} 
\end{equation}
The ${n}$-$th$ Minkowski distance $m_{h_i}^n$ in $\mathcal M_{h_i}$ is obtained as
\begin{align}
        m_{h_i}^n&=
        (|Lon_{h_i}^{n+1}-Lon_{h_i}^n|^p+|Lat_{h_i}^{n+1}-Lat_{h_i}^n|^p\nonumber\\
        &+|Alt_{h_i}^{n+1}-Alt_{h_i}^n|^p)^{1/p},
        \label{eq30}
\end{align}
where $p$ is the order of the Minkowski distances.
\par After receiving the $k$-$th$ packet from $h_i$, the central UAV calculates the Minkowski distance $m_{h_i}^{k-1}$ between $I_{h_i}^k$ and $I_{h_i}^{k-1}$. If
\begin{equation}
        m_{h_i}^{k-1}<\min \{m_{h_i}^1,m_{h_i}^2,...,m_{h_i}^n,...,m_{h_i}^{N}\},
\end{equation} 
the central UAV records $m_{h_i}^{k-1}$ in $\mathcal M_{h_i}$, replaces the $\max \{m_{h_i}^1,m_{h_i}^2,...,m_{h_i}^n,...,m_{h_i}^{N}\}$, and abandons the $k$-$th$ packet.
\subsubsection{Mechanism of Supplement}
If 
\begin{equation}
m_{h_i}^{k-1}\geq\max \{m_{h_i}^1,m_{h_i}^2,...,m_{h_i}^n,...,m_{h_i}^{N}\},
\end{equation}
 the central UAV records $m_{h_i}^{k-1}$ in $\mathcal M_{h_i}$, substitutes the $\min \{m_{h_i}^1,m_{h_i}^2,...,m_{h_i}^n,...,m_{h_i}^{N}\}$, and generates the packet supplement.

\begin{algorithm}[t]
        \caption{On-board processing algorithm}
        \begin{algorithmic}[1]
                \renewcommand{\algorithmicrequire}{\textbf{Input:}}
                \renewcommand{\algorithmicensure}{\textbf{Output:}}
                \REQUIRE{$N$ initial position vectors $I_{h_i}$, $I_{h_i}^k$, $I_{h_i}^{k-1}$, $I_{h_i}^{k-2}$, $I_{h_i}^{k-3}$. }
                \ENSURE{$x_s$, $y_s$, and $z_s$.}
                \STATE $\mathcal M_{h_i}\gets\{0\}$.
                \STATE $n\gets1$.
                \WHILE {$ n\leq N$}
                  \STATE Update $m_{h_i}^n$ based on Eq. (\ref{eq27}) and (\ref{eq30}).
                  \STATE Update $\mathcal M_{h_i}$ based on Eq. (\ref{eq29}).
                  \STATE $n\gets n+1$.
                \ENDWHILE
                \STATE Calculate $m_{h_i}^{k-1}$ by putting $I_{h_i}^k$ and $I_{h_i}^{k-1}$ into Eq. (\ref{eq27}) and (\ref{eq30}).
                \IF{$m_{h_i}^{k-1} < \min \{m_{h_i}^1,m_{h_i}^2,...,m_{h_i}^n,...,m_{h_i}^{N}\}$}
                \STATE Abandon $I_{h_i}^k$.
                \STATE $\max \{m_{h_i}^1,m_{h_i}^2,...,m_{h_i}^n,...,m_{h_i}^{N}\} \gets m_{h_i}^{k-1}$.
                \ENDIF
                \IF{$m_{h_i}^{k-1} \geq  \max \{m_{h_i}^1,m_{h_i}^2,...,m_{h_i}^n,...,m_{h_i}^{N}\}$}
                \STATE Calculate $D$ by Eq. (\ref{eq33}).
                \STATE Calculate $L$ by Eq. (\ref{eq34}).
                \STATE Calculate $x_s$, $y_s$ and $z_s$ by Eq. (\ref{eq35}), (\ref{eq36}), (\ref{eq37}), (38).
                \STATE Replenish $I_{h_i}^s$.
                \STATE $\min \{m_{h_i}^1,m_{h_i}^2,...,m_{h_i}^n,...,m_{h_i}^{N}\} \gets m_{h_i}^{k-1}$.
                \ENDIF
            \end{algorithmic}
            \label{A1}
 \end{algorithm}

 \begin{figure}[t]
        \centerline{\includegraphics[width=0.75\linewidth]{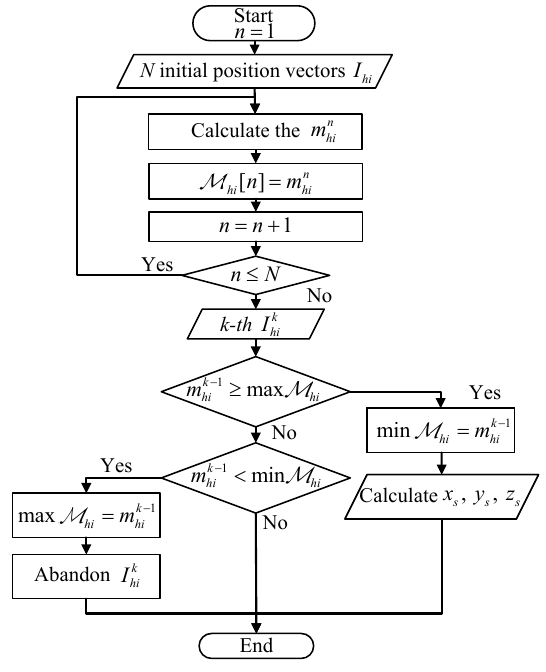}}
        \caption{On-board processing mechanism.}
        \label{f7}
\end{figure}
\par As shown in Fig. \ref{f6}, we utilize $I_{h_i}^k$, $I_{h_i}^{k-1}$, $I_{h_i}^{k-2}$ and $I_{h_i}^{k-3}$ to find the circumscribed sphere. The supplementary position information is at the intersection of the sphere and the straight line which passes the center and midpoint between $I_{h_i}^k$ and $I_{h_i}^{k-1}$, shown as the red cube in Fig. \ref{f6}. Besides, the center is ($x_o$, $y_o$, $z_o$). According to Cramer's Rule\cite{ref41}, the coefficient determinant $X =|\chi _1\ \chi _2\ \chi _3|$ is calculated as
\begin{equation}
        \begin{vmatrix} 
        Lon_{h_i}^k-Lon_{h_i}^{k-1} & Lat_{h_i}^k-Lat_{h_i}^{k-1}&Alt_{h_i}^k-Alt_{h_i}^{k-1} \\
        Lon_{h_i}^{k-2}-Lon_{h_i}^{k-3} & Lat_{h_i}^{k-2}-Lat_{h_i}^{k-3} &Alt_{h_i}^{k-2}-Alt_{h_i}^{k-3}\\
        Lon_{h_i}^{k-1}-Lon_{h_i}^{k-2} & Lat_{h_i}^{k-1}-Lat_{h_i}^{k-2} &Alt_{h_i}^{k-1}-Alt_{h_i}^{k-2}
        \end{vmatrix}.
        \label{eq33}
\end{equation}
The constant determinant $|B|$ is calculated as
\begin{equation}\label{eq34}
        |B|=\begin{vmatrix} 
                \beta_1\\
                \beta_2\\
                \beta_3  
        \end{vmatrix},
\end{equation}
which is further deduced as
\begin{equation}
        \begin{cases}
        \beta_1&=\{[(Lon_{h_i}^k)^2-(Lon_{h_i}^{k-1})^2+(Lat_{h_i}^k)^2-\\
        &(Lat_{h_i}^{k-1})^2+(Alt_{h_i}^k)^2-(Alt_{h_i}^{k-1})^2]\}/2,\\
        \beta_2&=\{[(Lon_{h_i}^{k-2})^2-(Lon_{h_i}^{k-3})^2+(Lat_{h_i}^{k-2})^2-\\
        &(Lat_{h_i}^{k-3})^2+(Alt_{h_i}^{k-2})^2-(Alt_{h_i}^{k-3})^2]\}/2,\\
        \beta_3&=\{[(Lon_{h_i}^{k-1})^2-(Lon_{h_i}^{k-2})^2+(Lat_{h_i}^{k-1})^2-\\
        &(Lat_{h_i}^{k-2})^2+(Alt_{h_i}^{k-1})^2-(Alt_{h_i}^{k-2})^2]\}/2.
        \end{cases}
\end{equation}
Moreover, the coordinate of center is deduced as
\begin{align}
    x_o=\frac{|B\ \chi _2\ \chi _3|}{X},\ 
    y_o=\frac{|\chi _1\ B\ \chi _3|}{X},\
    z_o=\frac{|\chi _1\ \chi _2\ B|}{X},\ 
    \label{eq35}
\end{align}
and the radius $r$ is 
\begin{equation}
r=\sqrt{(x_o-Lon_{h_i}^k)^2+(y_o-Lat_{h_i}^k)^2+(z_o-Alt_{h_i}^k)^2}.
\label{eq36}
\end{equation}
The position information in the supplementary packet between $I_{h_i}^k$ and $I_{h_i}^{k-1}$ is obtained by combining the following two equations:
\begin{equation}
        (x_s-x_o)^2+(y_s-y_o)^2+(z_s-z_o)^2=r^2,
        \label{eq37}
\end{equation}
and
\begin{align}
        &\frac{x_s-x_o}{(Lon_{h_i}^k+Lon_{h_i}^{k-1})/2-x_o}\nonumber\\
       =&\frac{y_s-y_o}{(Lat_{h_i}^k+Lat_{h_i}^{k-1})/2-y_o}\\
       =&\frac{z_s-z_o}{(Alt_{h_i}^k+Alt_{h_i}^{k-1})/2-z_o}.\nonumber         
\end{align}

\par In detail, the on-board processing mechanisms are further described in Algorithm 1, and the corresponding flowchart is shown as Fig.\ref {f7}.

\section{Simulation Results and analysis}\label{S6}
\begin{figure}[t]
        \centerline{\includegraphics[width=0.7\linewidth]{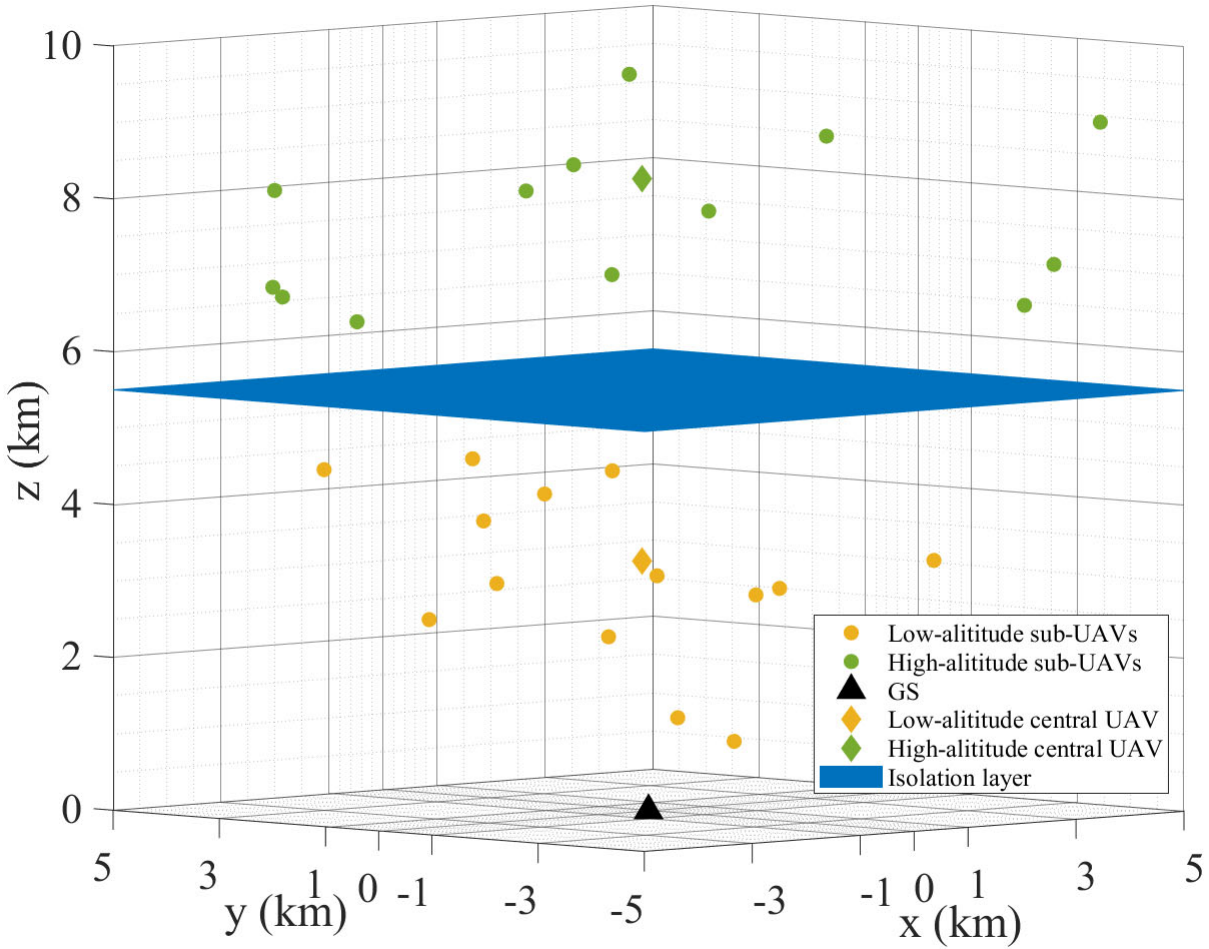}}
        \caption{Simulation scenario.}
        \label{f8}
\end{figure}

To evaluate the detailed performance, MATLAB is employed to simulate the 3D PPP distribution scenario of the sub-UAVs and central UAVs, as shown in Fig. \ref {f8}. The green icons representing UAVs in high-altitude airspace and yellow icons denoting UAVs in low-altitude airspace, are separated by the blue isolation layer, whose thickness $H_0$ is 1km. The circle, diamond, and triangle symbolize the sub-UAV, central UAV, and GS, respectively. The thickness of both airspace $\Delta H$ is fixed as 4.5km. Besides, the height of GS is set as 50m. The working frequency of the 5G system and ADS-B system is set as 3.5GHz and 1090MHz, respectively\cite{ref42}. The wavelength $\lambda_{5G} $ is 0.0857m and $\lambda_{ADS-B}$ is 0.2752m, according to $\lambda$=$c/f$. Moreover, the bandwidth $B_{5G}$ and $B_{ADS-B}$ are set as 100MHz and 1MHz, respectively\cite{ref6}. The transmitting power of the central UAVs are fixed as 20W while the transmitting power of the sub-UAVs varies in [1W, 20W]. In the wireless communication system, the value of Rice factor is usually determined according to the actual situation. Supposing fixing SNR, a larger Rice factor contributes to a smaller bit error rate, and a better system performance\cite{ref43}. Therefore, we take the Rice factor as 3 in the A2G channel, which comprehensively simulates the influence of reflected signals. Other parameters in simulations are listed in TABLE \ref {tab2}.

\begin{table}[t]
        \centering
        \caption{Key Parameters in the Simulations}
        \begin{center}
        \begin{tabular}{|p{1.5cm}<{\centering}|p{1.7cm}<{\centering}|p{1.5cm}<{\centering}|p{1.7cm}<{\centering}|}
        \hline  
        Parameter &Value&Parameter&Value\\
        \hline
        $f_{5G}$&3.5GHz&$f_{ADS-B}$&1090MHz\\
        \hline
        $\lambda_{5G}$& 0.0857m &$\lambda_{ADS-B}$&0.2752m\\
        \hline
        $B_{5G}$& 100MHz&$B_{ADS-B}$&1MHz\\
        \hline
        $n_0$& -174dBm/Hz&$\delta$&  2$\sim $4.9\\
        \hline
        $P_s$& 1W$\sim $20W&$\sigma$&5$\times10^3$\\
        \hline
        $P_{ch}$& 20W&$P_{cl}$&20W \\
        \hline
        $G_g$& 20dBi&$G_a$&23dBi\\
        \hline
        $\theta_h$& -14dB$\sim $-7dB&$\theta_l$&-14dB$\sim $-7dB\\
        \hline
        $λ\lambda_l $& 1$\sim $60&$λ\lambda_h $&1$\sim $60\\
        \hline
        $\Delta H$ & 4.5km&$H_0$&1km\\
        \hline
        $H_G$ & 50m &$\varepsilon_r$&15\\
        \hline
        \end{tabular}
        \label{tab2}
        \end{center}
\end{table}

\subsection{Performance Analysis of A2G channel}
\par We verify the network performance of the A2G channel by Eq. (\ref{eq10}) and (\ref{eq11}), which pertinently describe the process of signal transmission \cite{ref44}. The related indexes can be further subdivided into $PL_l$=$-10{\rm log}(P_G/P_{cl})$, $PL_h$=$-10{\rm log}(P_G/P_{ch})$, $\varpi_l$ and $\varpi_h$.

\par
\begin{figure}[t]
        \centerline{\includegraphics[width=0.8\linewidth]{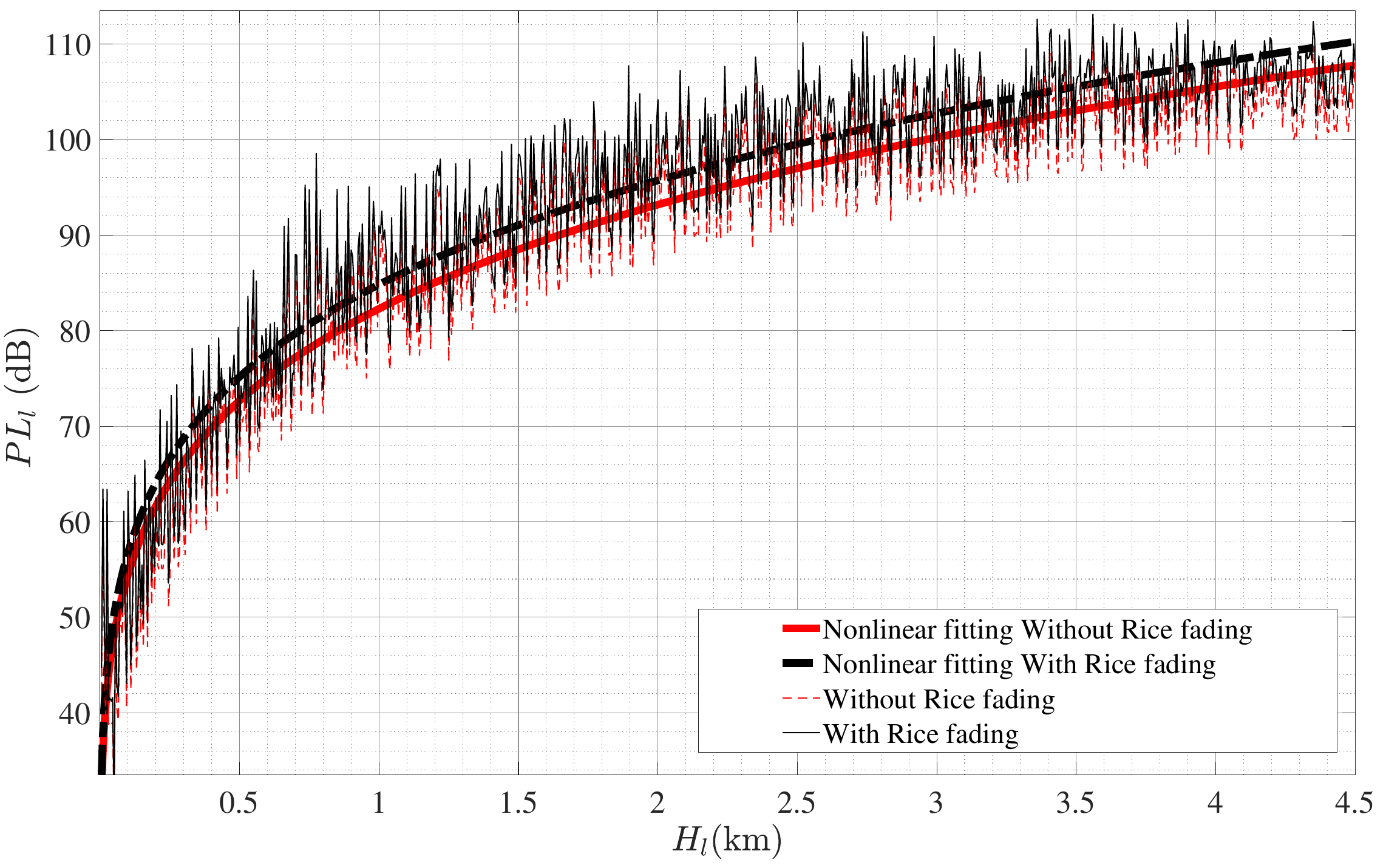}}
        \caption{Relationship between the $H_l$ and $PL_l$.}
        \label{f9}
\end{figure}

\begin{figure}[t]
        \centerline{\includegraphics[width=0.8\linewidth]{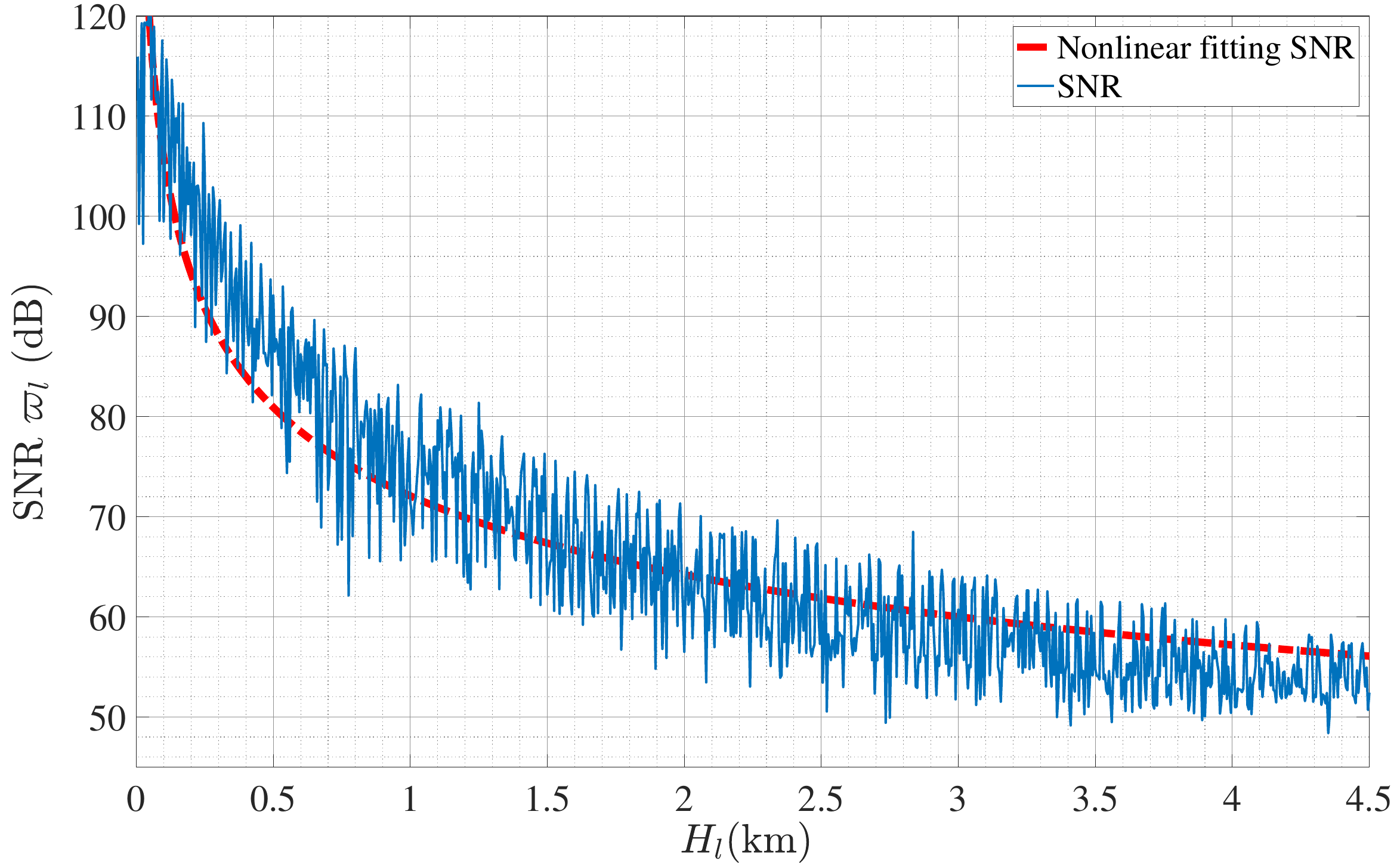}}
        \caption{Relationship between the $H_l$ and $\varpi_l$.}
        \label{f10}
\end{figure}

 Fig. \ref {f9} and Fig. \ref {f10} depict the A2G channel performance for the low-altitude central UAV $l_0$. In Fig. \ref {f9}, the black solid line indicates the path loss without Rice fading while the red dotted line demonstrates the path loss with Rice fading. The thick lines are corresponding nonlinear fitting curves. From Fig. \ref {f9}, the A2G models are oscillation models, following the law of free space path loss. With the height $H_l$ of the central UAV $l_0$ increasing, the path loss $PL_l$ also increases. After the introduction of Rice fading, $PL_l$ is growing, but the increment is not obvious in the low-altitude airspace. When $H_l$ is greater than 4km, the $PL_l$ curve tends to be smooth. In Fig. \ref {f10}, the blue solid line indicates the SNR of signal from $l_0$ to GS and the red dotted line is the corresponding nonlinear fitting. With the height $H_l$ of the central UAV $l_0$ increasing, the SNR $\varpi_l$ decreases. However, the decay speed of $\varpi_l$ gradually lowers with the addition of $H_l$. The curve tends to be flat when $H_l$ is higher than 4km in the 5G scenario. When $H_l$ approaches 0, the value of $PL_l$ changes abruptly. This is accounted that when $H_l$ is much less than the horizontal distance between $l_0$ and GS, the multi-rays transmitting model degenerates into a single-ray transmitting model. The height of the central UAV $l_0$ is recommended to be limited in respect to the practical receiving threshold of GS.
\begin{figure}[t]
        \centerline{\includegraphics[width=0.8\linewidth]{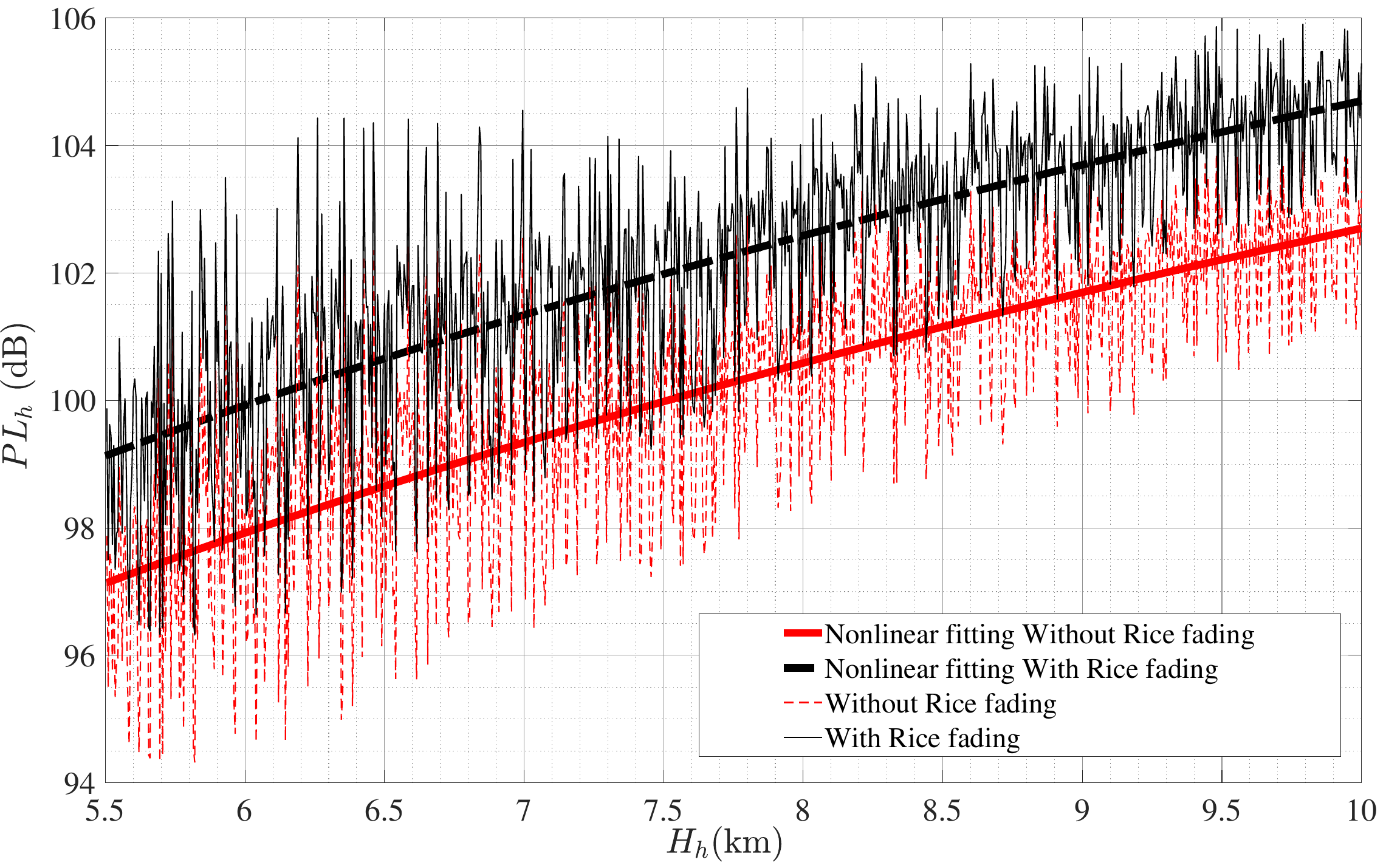}}
        \caption{Relationship between the $H_h$ and $PL_h$.}
        \label{f11}
\end{figure}
\begin{figure}[t]
        \centerline{\includegraphics[width=0.8\linewidth]{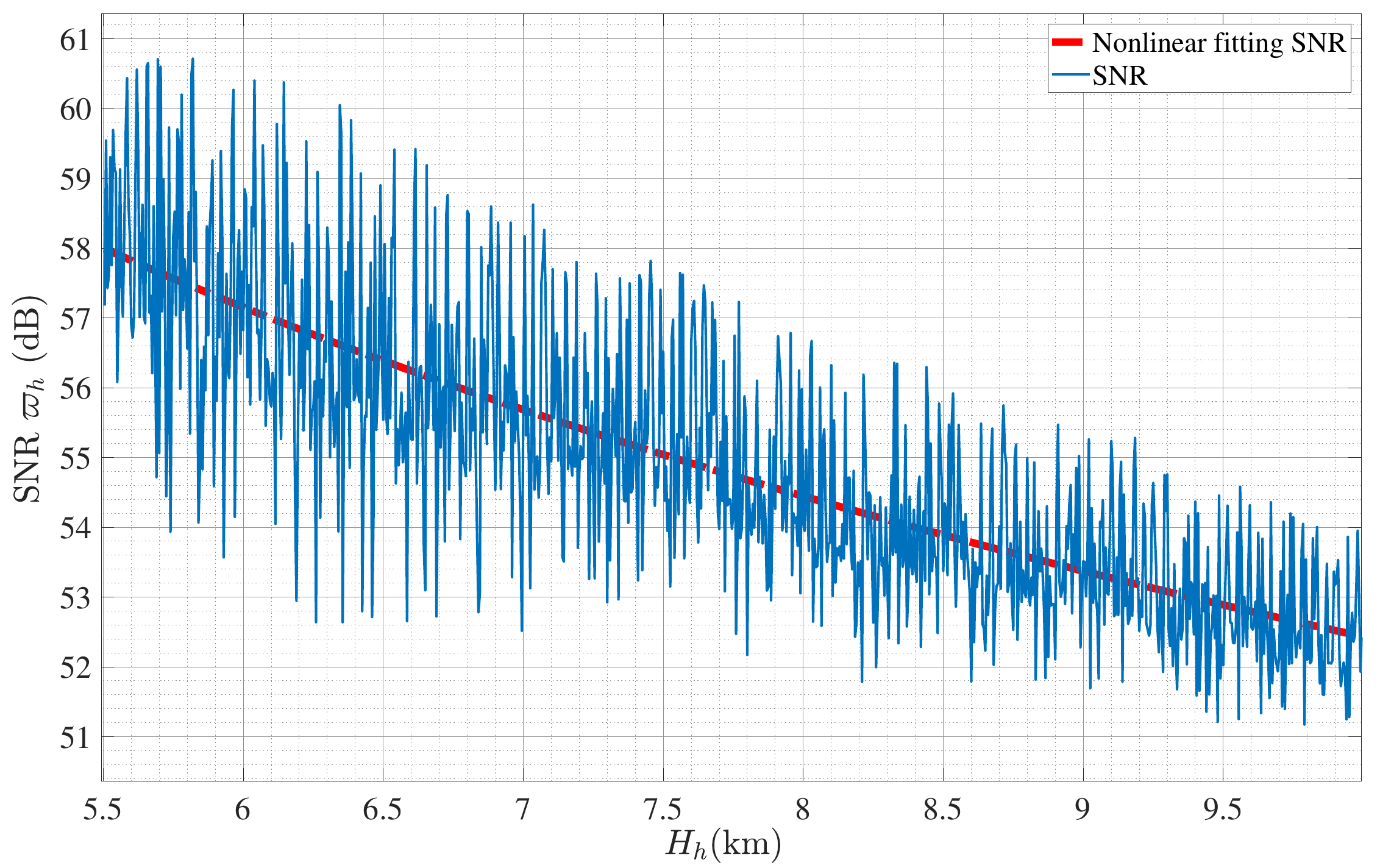}}
        \caption{Relationship between the $H_h$ and $\varpi_h$.}
        \label{f12}
\end{figure}
\par Fig. \ref {f11} and Fig. \ref {f12} illustrate the A2G channel performance for the high-altitude central UAV $h_0$. In Fig. \ref {f11}, the black solid line represents the path loss without Rice fading while the red dotted line symbolizes the path loss with Rice fading. The thick lines are corresponding nonlinear fitting curves. It is observed from Fig. \ref {f11} that $PL_h$ magnifies as $H_l$ enlarges, but the increment is not obvious in the high-altitude airspace in respect to the ADS-B central UAV $h_0$. Further, the difference between with and without Rice fading is obvious in the high-altitude airspace. Since we limit the height of the central UAV $h_0$ within (5.5km, 10km), the growth of $PL_h$ does not exhibit a smooth trend. In Fig. \ref {f12}, the blue solid line expresses the SNR of signal from $h_0$ to GS and the red dotted line is the corresponding nonlinear fitting. With the height $H_h$ of the central UAV $h_0$ rising, the SNR $\varpi_h$ reduces. Similarly, the decrement is small in the ADS-B scenario. The height of the central UAV $h_0$ should be determined according to the actual requirement of GS.

\begin{figure*}[t] 
        \centering
       \begin{minipage}{1\linewidth}
      \centering
        \subfloat[Relationship between the $\lambda$ and $\gamma$.]{
        \includegraphics[width=0.325\linewidth]{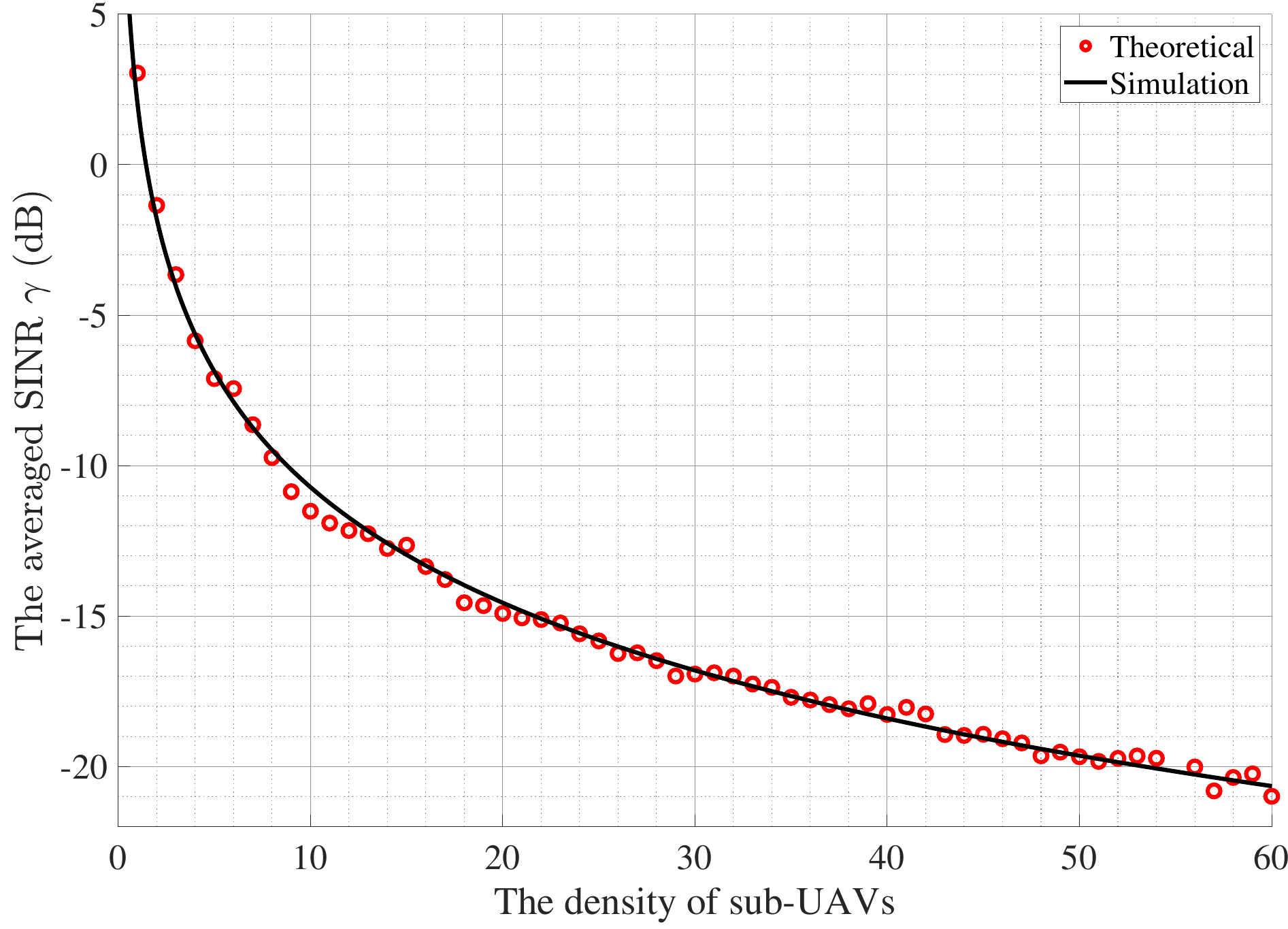}
       \label{f13.1}}
        \subfloat[Relationship between the $P_s$ and $P_{cov}$.]{
       \includegraphics[width=0.31\linewidth]{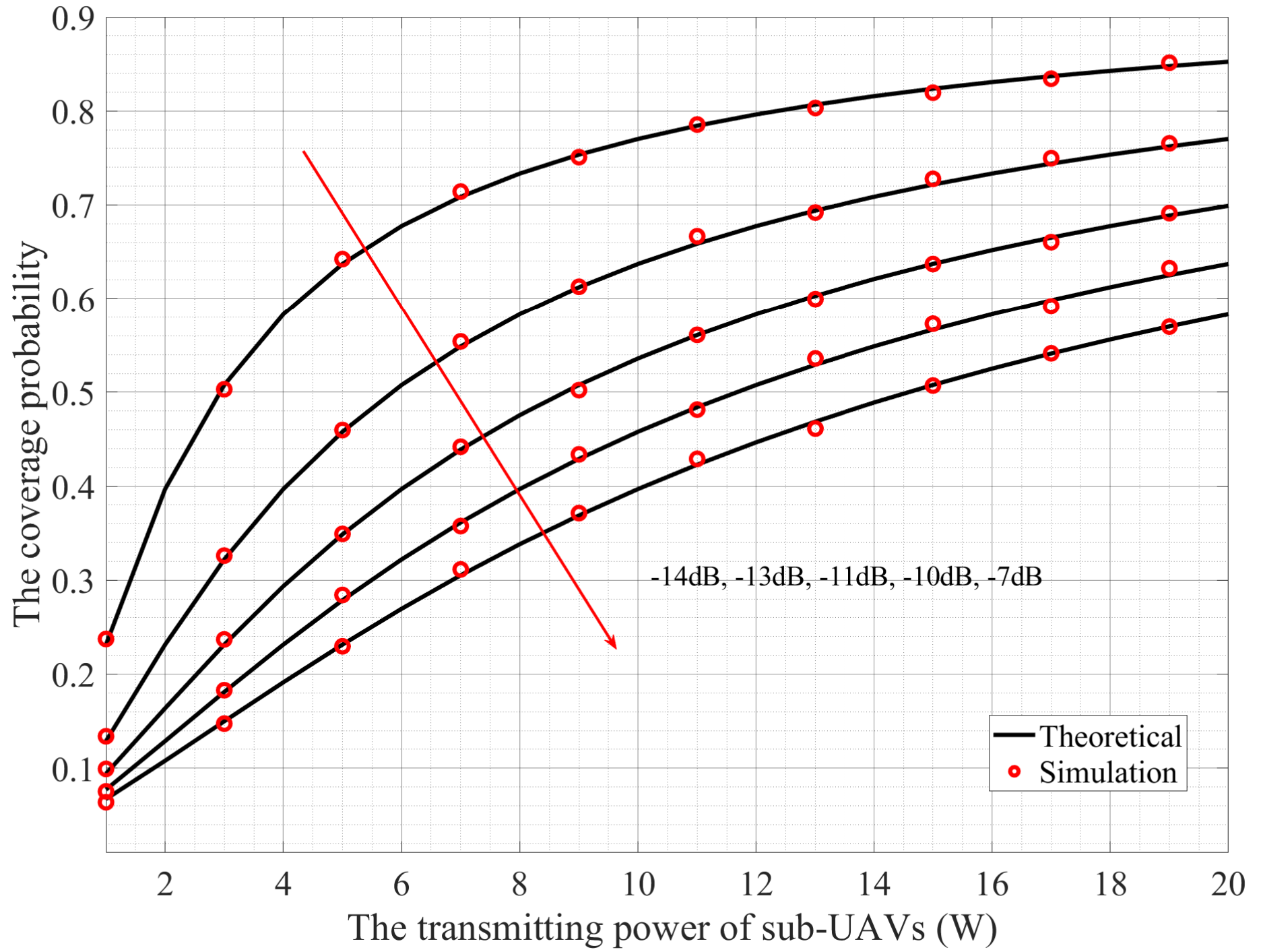}
        \label{f13.2}}
        \subfloat[Relationship between the $\delta$ and $P_{cov}$.]{
        \includegraphics[width=0.31\linewidth]{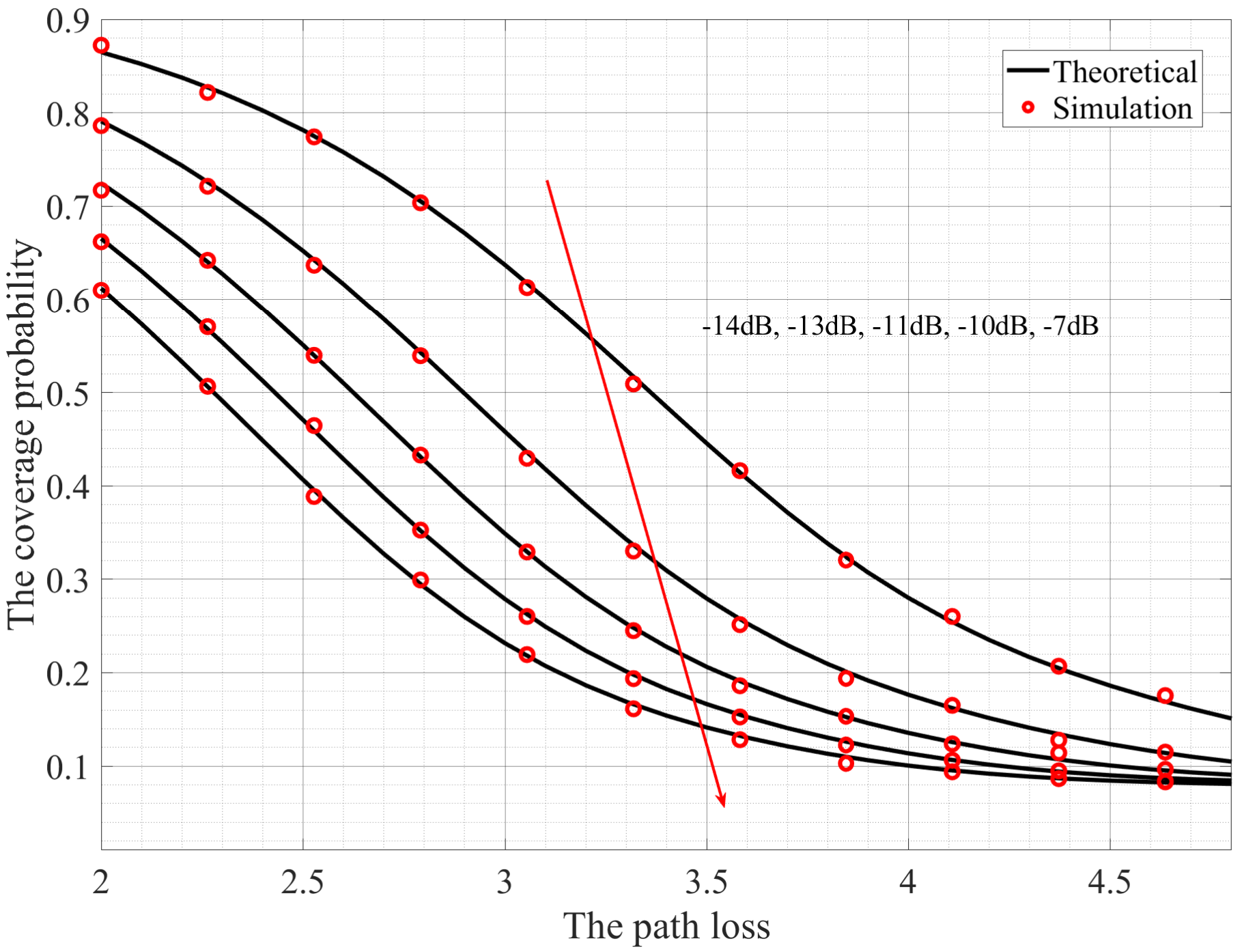}
       \label{f13.3}}
        \caption{Performance analyze of A2A.}
        \label{f13}
        \end{minipage}
\end{figure*}

\subsection{Performance Analysis of A2A Channel}
\par Fig. \ref {f13.1} shows the relationship between the density of sub-UAVs and averaged SINR of the signal received by the central UAV. Both airspaces have the same size, and all sub-UAVs send packets to the central UAVs, so the performance analysis of the A2A does not distinguish between the high-altitude and the low-altitude. As the density of sub-UAVs increases, the averaged SINR decreases. Besides, the simulation results coincide with the theoretical results, which validates the accuracy of theoretical analysis. The increment of sub-UAVs leads to abundant signals propagating in the airspace. Thus, there exist more interference signals when the central UAV receives a specific signal. When the density of sub-UAVs exceeds 40, the averaged SINR of the signals degenerates to around -20, which is harmful to signal processing. Therefore, the density of sub-UAVs should adapt to practical received threshold. 
\par Fig. \ref {f13.2} describes the relationship between the coverage probability and transmitting power of sub-UAVs under different received thresholds. Besides, the simulation results coincide with the theoretical results, which validates the accuracy of theoretical analysis. Moreover, the received threshold increases with the direction of the arrow. The path loss index and density are set as 2 and 20, respectively. With the power rising, the signal coverage probability grows. Besides, when the power is greater than 17W, it shows a smooth trend in the increment of the coverage probability regardless of thresholds. However, the enlargement of the threshold leads to the coverage probability falling. Supposing the threshold is -14dB, fixing the transmitting power as 7W makes the highest efficiency of energy investment, since the gain of the coverage probability from the increment of a unit transmitting power is continuously decreasing.
\par Fig. \ref {f13.3} presents the relationship between the coverage probability and path loss index under different received thresholds. The transmitting power and density are fixed as 20W and 20, respectively. Besides, the simulation results coincide with the theoretical results, which validates the accuracy of theoretical analysis. Moreover, the received threshold increases with the direction of the arrow. When the path loss index is less than 2.5, its effect on the decrement of the coverage probability is relatively small. However, when the path loss is greater than 2.5, the performance of the communication system deteriorates dramatically. When the path loss is greater than 4, only in the scenario of -14dB threshold, the coverage probability is still relatively considerable. However, in the other four cases, the coverage probability approaches 0.1. In the above four cases, the system performance can only be improved by lowering the threshold, intensifying the transmitting power or reducing the density of sub-UAVs.

\subsection{Experiments of On-board Processing Algorithm}
\par
\begin{figure}[t]
        \centerline{\includegraphics[width=0.85\linewidth]{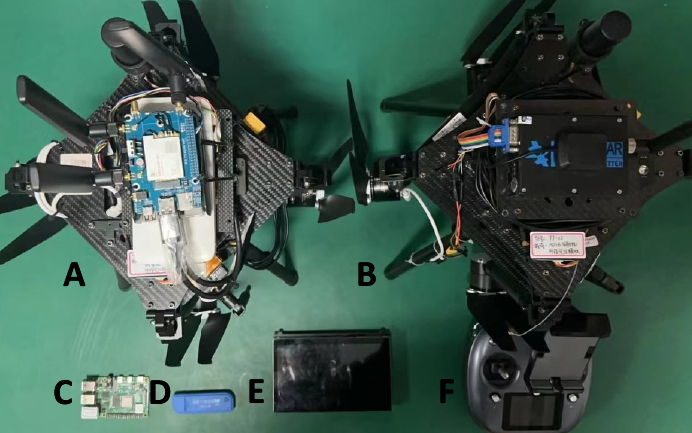}}
        \caption{Experimental equipments.}
        \label{f14}
\end{figure}

 As shown in Fig. \ref {f14}, we collect the actual flight data of the UAV for algorithm verification. \textbf{A} is the UAV equipped with 5G module (SIM8262E-M2), while \textbf{B} is the UAV equipped with ADS-B out device (PT050X). Additionally, \textbf{C} is the Raspberry PI (4-Model-B), playing the role of GS and processing the position packets. \textbf{D} is the Realtek software defined radio as the antenna to receive the position packets. Besides, \textbf{E} is the display and \textbf{F} is the UAV controller. The constructions of the sending and receiving terminals are built in our lab. Further, in the experiment, we consider A and B as sub-UAVs and C as the central UAV.

\begin{figure*}[t]
        \centering
        \subfloat[Original data \ding{172}.]
        {\includegraphics[width=0.40\textwidth]{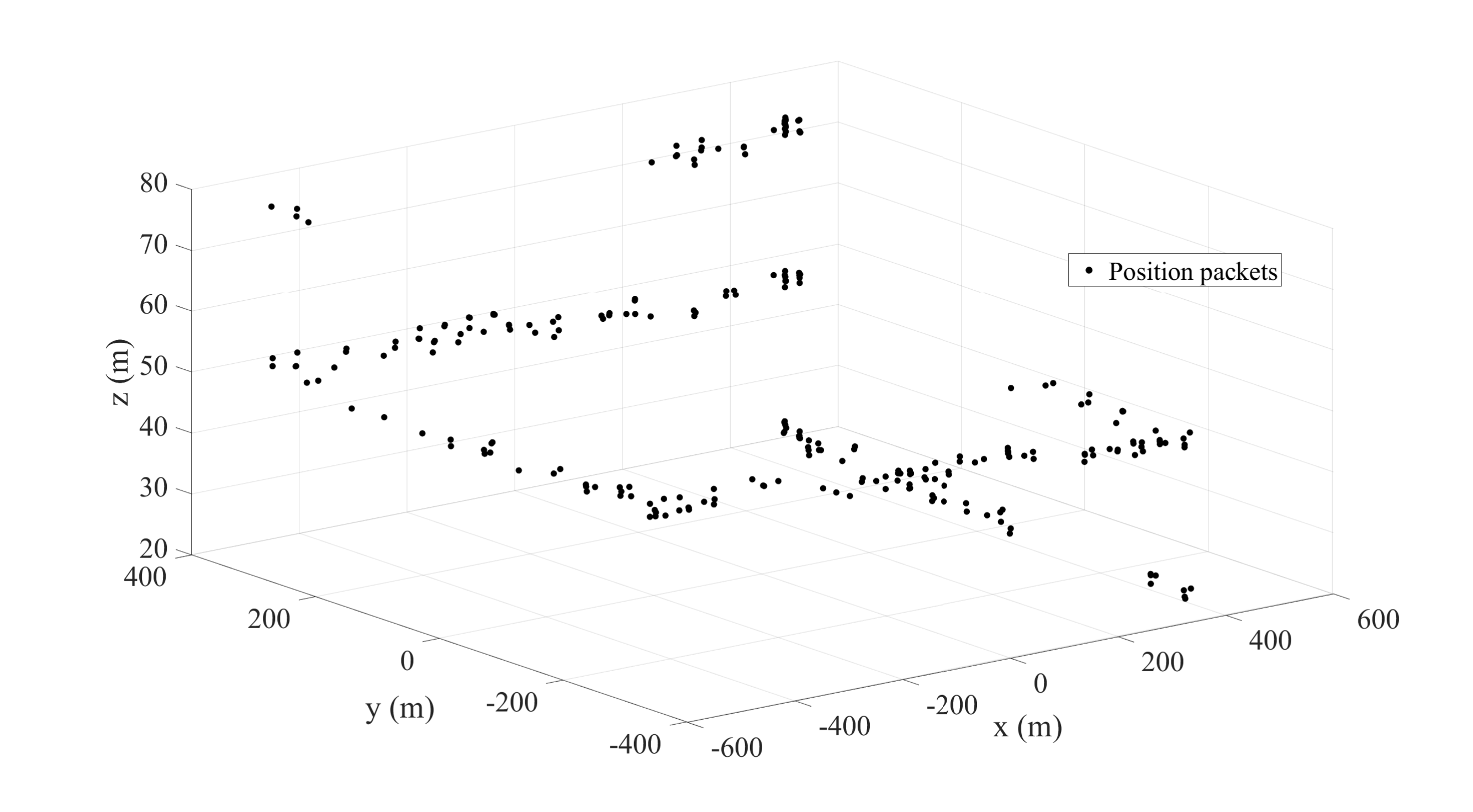}\label{f15a}}    
        \quad
        \subfloat[Optimized data \ding{172}.]      
        {\includegraphics[width=0.43\textwidth]{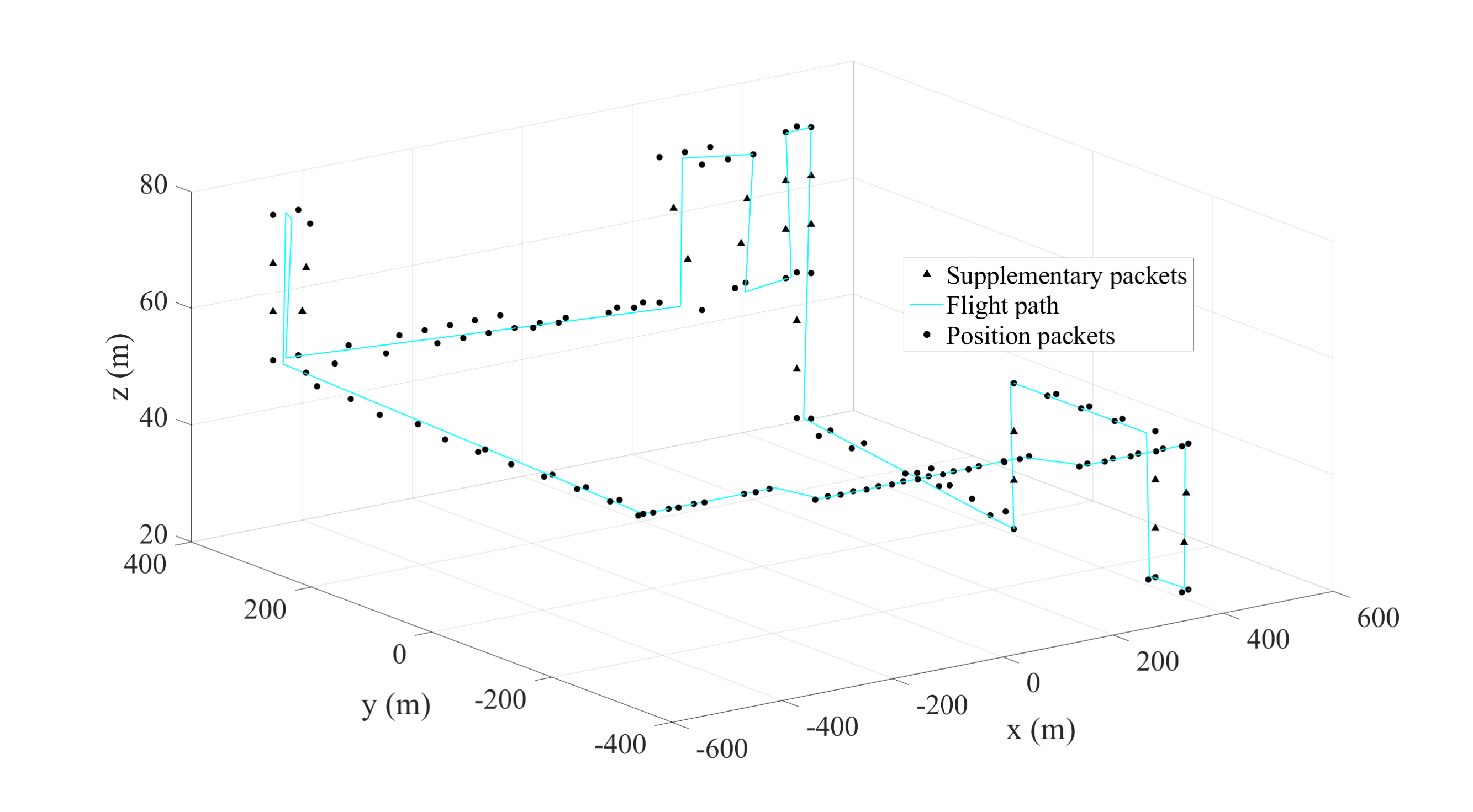}\label{f15b}}
        \quad
        \subfloat[Original data \ding{173}.]
        {\includegraphics[width=0.40\textwidth]{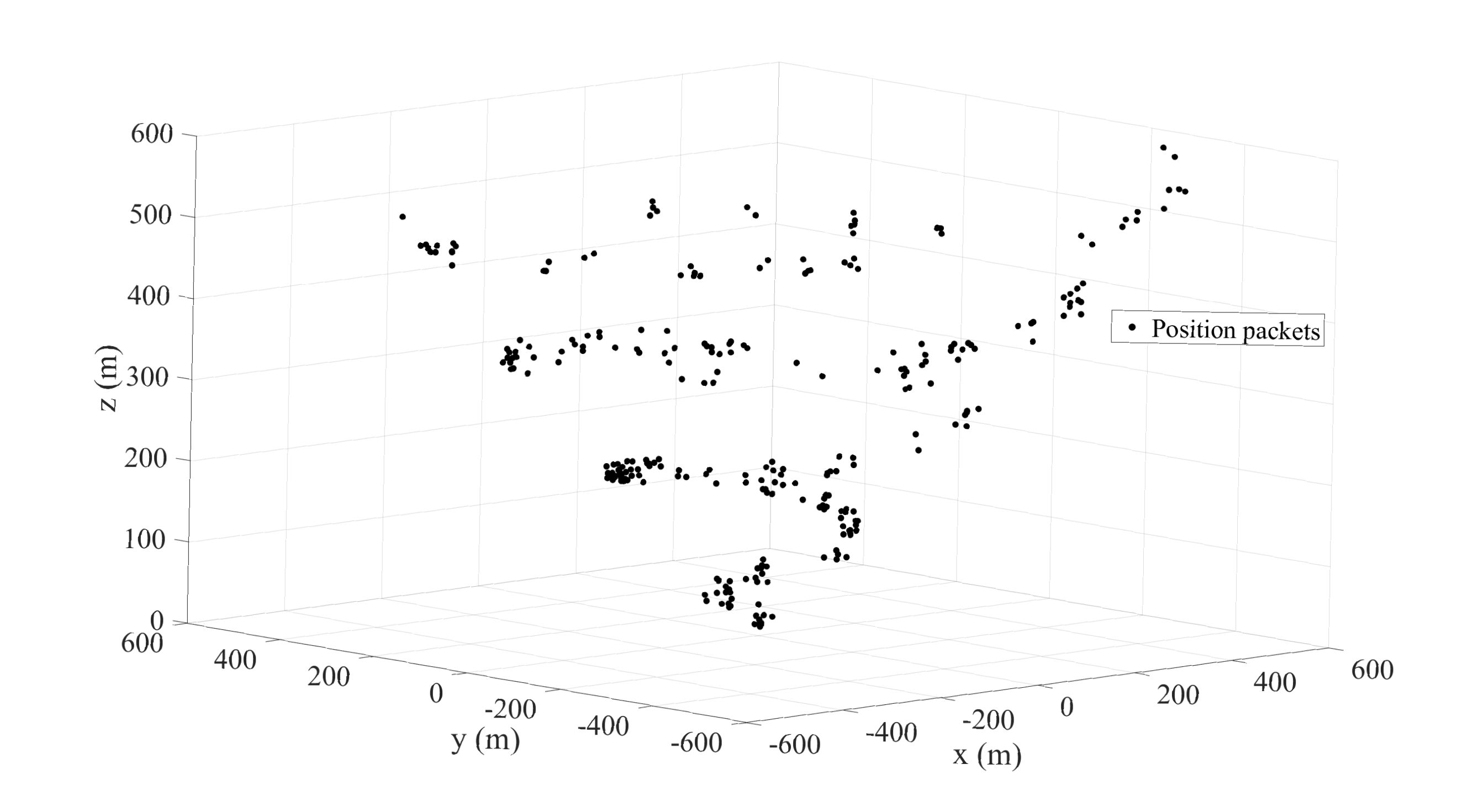}\label{f15c}}
        \quad
        \subfloat[Optimized data \ding{173}.]
        {\includegraphics[width=0.43\textwidth]{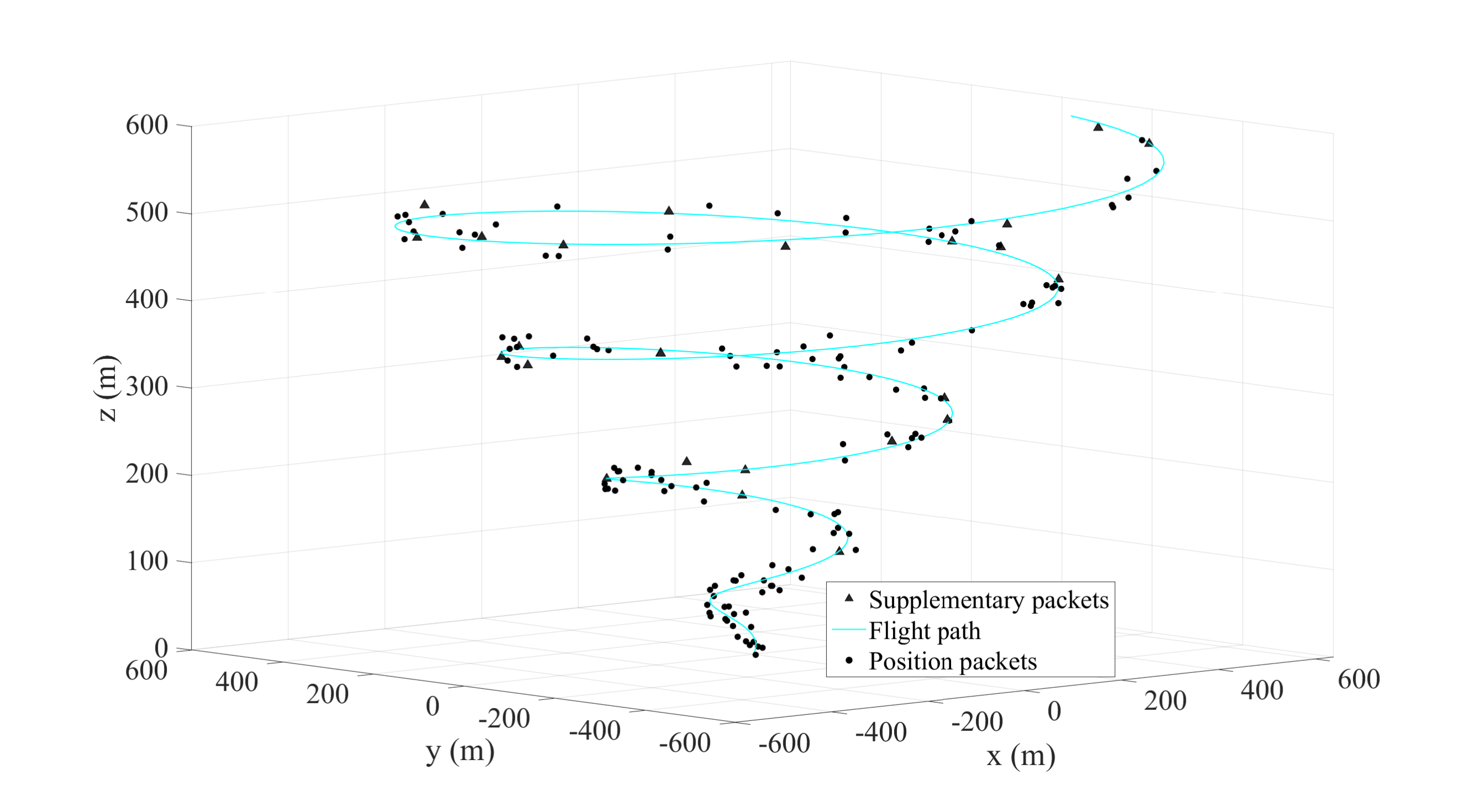}\label{f15d}}
        \quad
        \subfloat[Original data \ding{174}.]
        {\includegraphics[width=0.40\textwidth]{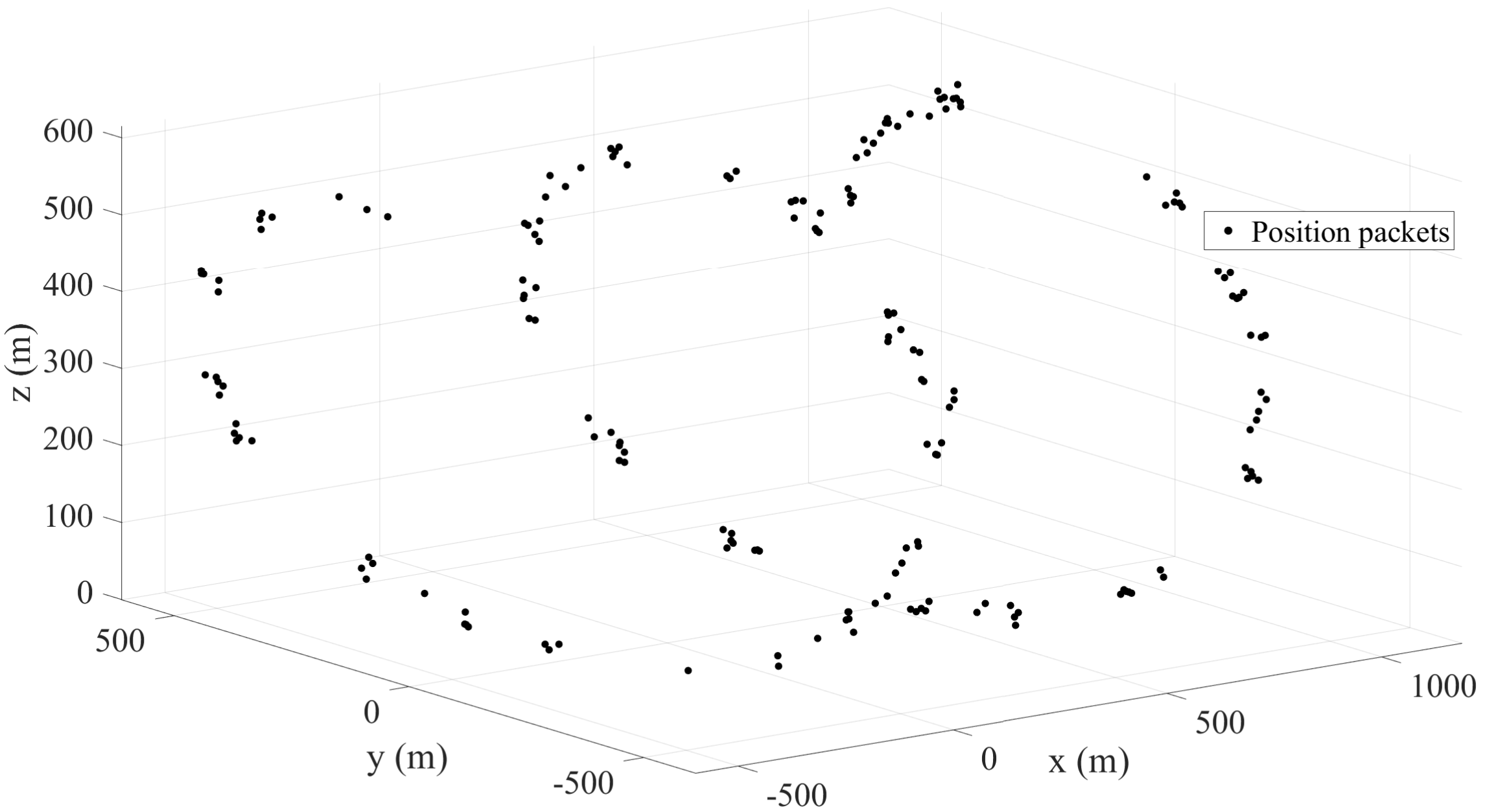}\label{f15e}}
        \quad
        \subfloat[Optimized data \ding{174}.]
        {\includegraphics[width=0.43\textwidth]{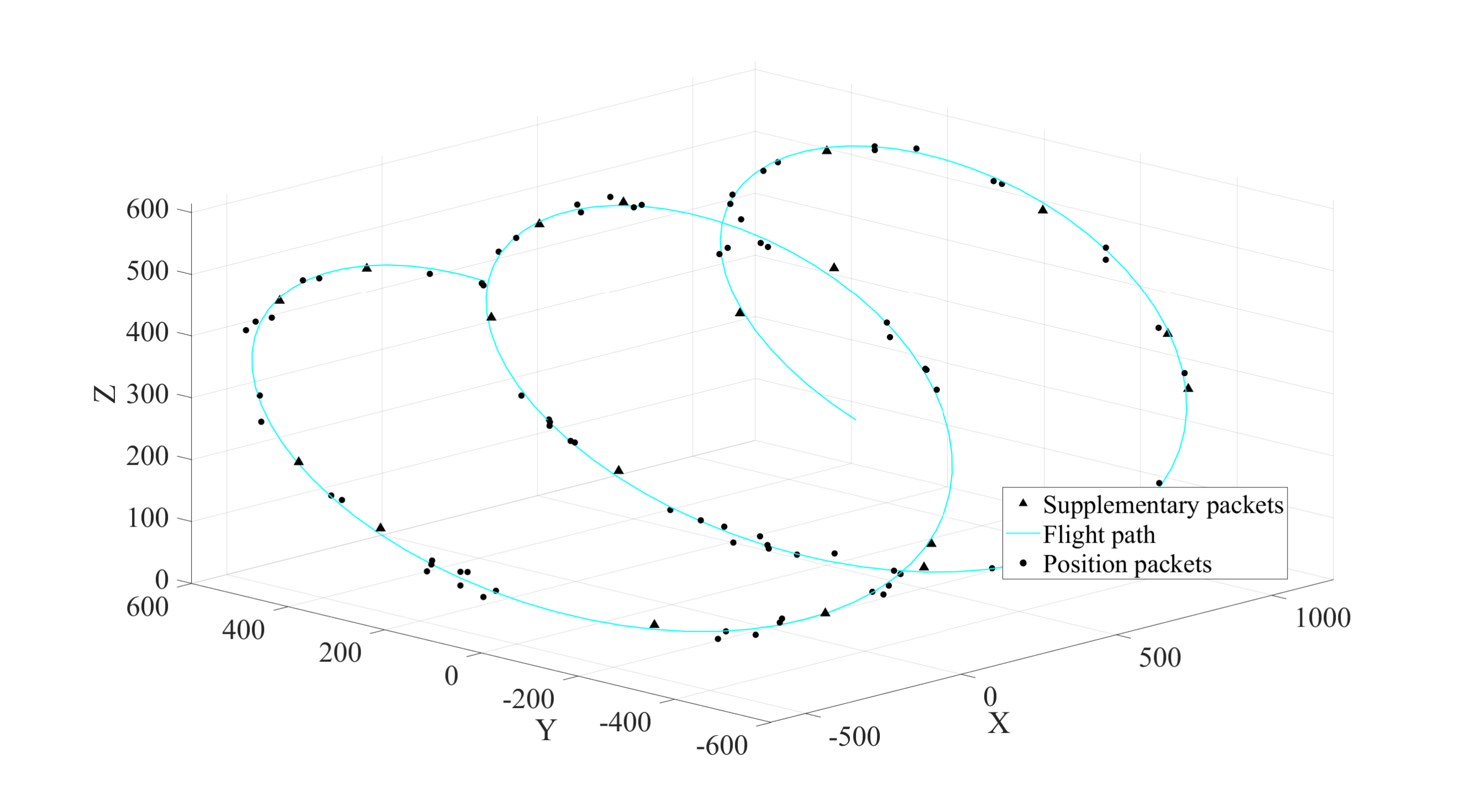}\label{f15f}}
        \quad
        \caption{On-board data processing of position packets.}
        \label{f15}
\end{figure*}

\par Fig. \ref {f15} demonstrates the on-board data processing of position packets. Based on the mechanism, the central UAV processes the position packets of the sub-UAVs accordingly after receiving, and then transmits the packets to GS. On the one hand, there exist random perturbations or similar information during positioning and transmission. Relaying these redundant information to GS aggravates the burden of data processing, which is time-consuming and energy-consuming. Therefore, we utilize the packet abandonment mechanism to reduce the redundant information. On the other hand, the packets transmission is affected by many random factors, leading to packet error or loss in serious cases. Further, the above situations trigger discontinuous trajectory, thus decreasing the observability. Therefore, we use the packet supplement mechanism to replenish a supplementary packet, achieving a smooth display of the UAV trajectory. Data \ding{172} is actual flight packets, but this scenario is not universal, since the altitude of the UAV does not change continuously. Therefore, we collect simulation data of \ding{173} and \ding{174} for additional verifications.
\par Fig. 15\subref {f15a} and Fig. 15\subref {f15b} show the original data and optimized data of trajectory \ding{172}. The dots represent the received position packets, and the triangles represent the supplementary packets. In this scenario, the redundant data is effectively reduced by 52.55\%, and 20 packets are added, accounting for 14.39\% of the optimized data. However, trajectory \ding{172} is a special case, since the altitude stays invariable for most of trajectories. Therefore, the spherical supplement mechanism is ineffective in this scenario. Instead, the weighted average is used for the supplement, which means we utilize adjacent points to perform the linear supplement.
\par Fig. 15\subref {f15c} and Fig. 15\subref {f15d} display the original data and optimized data of trajectory \ding{173}. In this scenario, the longitude, latitude and altitude of the UAV change over time, which is a more universal scene. Therefore, the spherical supplement mechanism is adopted. In trajectory \ding{173}, the redundant data is effectively reduced by 52.41\%, and 26 packets are added, accounting for 16.05\% of the optimized data.
\par Fig. 15\subref {f15e} and Fig. 15\subref {f15f} illustrate the original data and optimized data of trajectory \ding{174}. In this scenario, the longitude, latitude and altitude of the UAV also change over time. Therefore, the spherical supplement mechanism is adopted. In trajectory \ding{174}, the redundant data is effectively reduced by 55.49\%, and 19 packets are added, accounting for 18.81\% of the optimized data.
\par However, the algorithm sacrifices accuracies to some extent in two unusual cases. In particular, the first case is that the altitude of UAV remains unchanged for an extended period. Besides, the second case is that the velocity of UAV undergoes a transient change.

\vspace{-0.2cm}

\section{Conclusions}\label{S7}
In this paper, we design the cooperating framework of ADS-B in B5G for hierarchical UAV networks. Moreover, since the redundancy or trajectory information loss during transmission aggravates the performance and decreases the observability, the MEC based on-board processing algorithm is proposed. In detail, we build the system of the A2G and A2A channels based on the deterministic modeling and stochastic modeling, respectively. Besides, we derive the related analytic formulas, analyze the performance of the A2G and A2A systems, and provide the corresponding suggestions about the height, density and transmitting power of UAVs. Further, the algorithm is verified by simulations as well as experiments. The results indicate that the proposed mechanism effectively filters out redundant data with similar information and makes supplies for the discontinuities.

\par In furture works, we will conduct more real-world experiments. Moreover, we will build a multi-mode hybrid network for UAVs based on more access methods. Furthermore, considering the hybrid network, we will explore how to enable GS to dynamically combine the multi-source heterogeneous data, thus offering more solid assurance and robust support for low-altitude development.

\vfill
\end{CJK}
\end{document}